\documentclass[aps,pre,superscriptaddress,amssymb,floatfix,reprint,notitlepage]{revtex4-1}
\usepackage[centertags]{amsmath}
\usepackage{amsfonts}
\usepackage{verbatim}
\usepackage{amsthm}
\usepackage{graphicx}
\usepackage{bbm}

\newcommand{\figtext}[1]{\linespread{1.1} \small #1}

\newcommand{\fig}{Fig.\ }
\newcommand{\sect}{Sec.\ }
\newcommand{\eqn}{Eq.\ }
\newcommand{\eqns}{Eqs.\ }

\newcommand{\physrep}{Phys.\ Rep.\ }

\newcommand{\lmax}{{\ell_\text{max}}}
\newcommand{\q}{q}
\newcommand{\Q}{Q}
\newcommand{\jin}{J_\text{in}}
\newcommand{\jout}{J_\text{out}}
\newcommand{\rhobg}{\rho_{\text{bg}}}

\begin{document}


\title{Condensation transition and drifting condensates in the accelerated exclusion process}

\author{Ori Hirschberg}
\affiliation{Department of Physics, Technion, 3200003 Haifa, Israel}
\affiliation{Department of Physics of Complex Systems, Weizmann
Institute of Science, 76100 Rehovot, Israel}
\author{David Mukamel}
\affiliation{Department of Physics of Complex Systems, Weizmann
Institute of Science, 76100 Rehovot, Israel}

\date{\today}


\begin{abstract}
Recently, it was shown that spatial correlations may have a drastic
effect on the dynamics of real-space condensates in driven
mass-transport systems: in models with a spatially correlated steady
state, the condensate is quite generically found to drift with a
non-vanishing velocity. Here we examine the condensate dynamics in
the accelerate exclusion process (AEP), where spatial correlations
are present. This model is a ``facilitated'' generalization of the
totally asymmetric simple exclusion process (TASEP) where each
hopping particle may trigger another hopping event. Within a
mean-field approach that captures some of the effects of
correlations, we calculate the phase diagram of the AEP, analyze the
nature of the condensation transition, and show that the condensate
drifts, albeit with a velocity that vanishes in the thermodynamic
limit. Numerical simulations are consistent with the mean-field
phase diagram.
\end{abstract}

\maketitle

\section{Introduction}

A traffic jam forms on a highway; a macroscopically linked hub
develops in a complex network; marbles in a shaken compartmentalized
box cluster in a single compartment; a finite fraction of the
capital in a macroeconomic system is held in the hands of few
agents. These varied phenomena can all be described as types of
\emph{real-space condensation}, where a macroscopic fraction of some
``mass'' (cars, links, marbles, capital, etc.) is typically
concentrated in a microscopic part of the system
\cite{evanszrpreview,MajumdarCondensationLesHouches,Schadschneider2010Book,%
networksevolutionbook,*DorogovstevEtal2008NetworksRMP,vanderweele2001,%
BouchaudMezard2000WealthCondensation}. Condensation in such systems
usually sets in via a phase transition which is mathematically
similar to Bose-Einstein condensation: at low densities the system
is in a homogenous disordered phase, and when the density is
increased beyond a critical value a condensate forms. Such
condensation transitions occur both in and out of equilibrium, and
their static and dynamical properties have received much attention
in recent years
\cite{MajumdarEtalChipping1998,EvansZRPcondensation,kafrietal2002criterion,%
HanneyEvans2004ZRP2species,EvansEtAl2005MassTransportCondensation,%
EvansEtal2006PairFactorizedPRL,WaclawEtal2009PairFactorizedPRL,*WaclawEtal2009PairFactorizedJSTAT,%
GrosskinskySchutzZRP1stOrder,GrosskinskyEtal2011Inclusion,%
GodrecheLuck2012Inhomogeneous,Barma2013PRLintermittency,*Barma2013ASIPintermittency,%
JurajEtal2014ConstraintDriveCondensation,*JurajEtal2014ConstraintDriveCondensationLong,%
Marsili2013condensationFatTailed,*MarsiliEtal2014CondensationLDF,ChlebounGrosskinsky2014CondensationFactorized,%
ConcannonBlythe2014NonMarkovTASEP,BarMukamel2014MixedOrderPRL,*BarMukamel2014MixedOrderLong}.

The study of such condensation transitions is simplest in exactly
solvable models. The most notable examples of models of this type
are zero-range processes (ZRPs)
\cite{evanszrpreview,MajumdarCondensationLesHouches,Schadschneider2010Book}.
These are models of transport in which particles hop stochastically
among sites with rates that depend only on the occupation number of
the departure site. In this class of models, the steady-state
distribution factorizes into a product of single site terms and can
ba calculated exactly. Therefore, one may analyze the precise
conditions under which condensation takes place in a ZRP
\cite{EvansZRPcondensation}. ZRPs have been used to model
condensation in a variety of contexts, including the examples of
condensing systems listed in the opening lines of this paper
\cite{kaupuzsetal2005zrptraffic,%
BurdaEtal2001NetworksCondensation,*DorogovtsevEtal2003NetworksCondensation,%
*angeletal2005zrpnetworks,*AngelEtal2006zrpNetworks,%
torok2005granularzrp,BurdaEtal2002WealthZRP}.

In generic systems, the steady-state distribution is not known and
exact results about condensation are scarce. In some specific
models, the problem can by bypassed by specially tailored methods
which allow one to study condensation
\cite{MajumdarEtalChipping1998,RajeshChippingSym2001,RajeshChippingAsym2002}.
However, in most cases, exact methods for the analysis of
condensation are not available. In these cases, one usually resorts
to approximate methods such as mean-field (MF) approximations, where
correlations between sites are neglected. Although inexact, MF
methods often lead to a qualitative description of the collective
behavior of the model, and thus provide insight into the phenomena
under study. Usually, MF descriptions not only neglect correlations
among sites, but also assume that the system is homogeneous. Thus,
one studies condensation in the model under consideration by
effectively describing it as a ZRP.

One drawback of such MF methods is that they do not adequately
describe the condensed phase. This phase is not homogeneous, as the
formation of a condensate breaks translational invariance, and usual
MF treatments ignore such inhomogeneities. Recently, a MF method was
proposed which is better suited for the study of the inhomogeneity
of the condensed phase \cite{CondDriftPRE2013}. In a nutshell,
different sites are assumed in this MF scheme to be independent but
not necessarily identically distributed. The occupation probability
of a site is thus allowed to depend on its distance from the
condensate.

An important aspect of condensing systems which is modified by the
presence of spatial correlations concerns the dynamics of the
condensate. In the ZRP, the condensate location remains static for
long periods of time (the duration of which diverges with the system
size faster than quadratically), until a fluctuation eventually
leads the condensate to relocate \cite{godrecheluck2005condensate}.
When it relocates it does so to a random site. In Ref.\
\cite{CondDriftPRE2013} it was shown, using the aforementioned
modified MF method, that spatial correlations in the steady state
often modify this condensate dynamics: they may lead the condensate
to drift along the system with a non-zero velocity in any finite
system size. The dynamics of condensates is currently an active line
of study, as condensing systems provide one of the simplest settings
in which collective and emergent motion can be studied
\cite{godrecheluck2005condensate,HirschbergEtal2009,*HirschbergEtal2012Long,%
WaclawEvans2012MovingCondensate,*WaclawEvans2014MovingCondensateLong,%
beltramlandim2008,LandimMetastabilityAsymmetricZRP,%
ChlebounGrosskinsky2014DynamicalMetastability,WhitehouseEvansEtal2014MovingCondensate,%
ArmendarizEtal2015MetastabilityZRP,ChauEtal2015ExplosiveCondSymmetric}

In the present paper, we demonstrate the use of the MF method of
Ref.\ \cite{CondDriftPRE2013} by applying it to the study of
condensation in a recently introduced accelerated exclusion process
(AEP) \cite{DongAEPprl2012}. This model is a ``facilitated'' version
of the well known totally asymmetric simple exclusion process
(TASEP), in which a particle hop may trigger a second (simultaneous)
hop, see below. Numerical and analytical studies of finite AEP
systems and of some specific infinite-system limits have suggested
that a condensation phase transition occurs in the model. However,
lacking an analytical description of the steady state, the
conditions for the occurrence of this apparent phase transition in
the thermodynamic limit were not known. A subsequent mean-field
study of the model was useful in clarifying its behavior in the
homogeneous disordered phase, but did not resolve the questions
regarding the phase transition \cite{DongAEPpre2013} (see Refs.\
\cite{Merikoski2013SymmetricAEP,BhatKrapivsky2014AEPavalanches} for
studies of related models). Here, using the MF scheme of Ref.\
\cite{CondDriftPRE2013}, we calculate the MF phase diagram of the
model and explain some of the numerical findings of Ref.\
\cite{DongAEPprl2012}. Furthermore, Our analysis suggests that the
model has a drifting condensate, and thus the AEP provides another
example for the condensate-drift mechanism studied in Ref.\
\cite{CondDriftPRE2013}.

The paper is organized as follows. We begin in \sect
\ref{sec:ModelDescription} by defining the model. Two
representations are presented: the original definition of Ref.\
\cite{DongAEPprl2012}, and an alternative description in which
similarities of the model to a ZRP are more apparent. After
summarizing the main results of the paper in \sect
\ref{sec:MainResults}, the analysis of the model is presented in
\sect \ref{sec:Calculation}, where we calculate its MF phase diagram
and discuss the condensate dynamics. Predictions derived from our MF
analysis are shown in \sect \ref{sec:numerical} to agree
qualitatively, and in some instances also quantitatively, with with
results of numerical simulations. \sect \ref{sec:Conclusion}
summarizes the conclusions of our analysis. Some of the more
technical aspects of the analysis and numerics are presented in the
Appendices.

\section{Description of the model and mapping to a
ZRP}\label{sec:ModelDescription}

The AEP dynamics can be represented in two seemingly different ways
which are in fact equivalent. One representation is as a variant of
the TASEP and the other as a variant of the ZRP. The equivalence of
these two descriptions corresponds to the well known mapping of
exclusion processes to ZRPs \cite{evanszrpreview}. In this section
we present the definition of the model in these two different
pictures, and then discuss how the mapping from one picture into the
other is achieved. Below we refer to the model in the exclusion
process picture simply as the AEP, while we call the other ``the ZRP
picture''. Note that we use below the term ``ZRP picture'' to
highlight similarities to the ZRP, even though the model is not
strictly ``zero range'' in nature.

\begin{figure}
  \center
  \includegraphics[width = 0.45\textwidth]{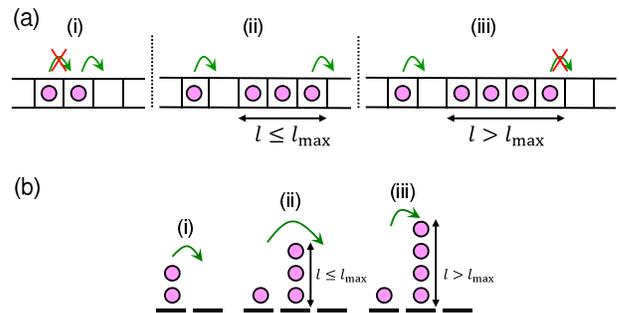}
  \caption{\label{fig:aepscheme}\figtext{(a) A schematic representation
  of the dynamics of the AEP for $l_{\max} = 3$. (i) Particles hop in a totally asymmetric fashion
  subject to exclusion interactions. (ii) When a particle joins a cluster of
  length $l\leq l_{\max}$, it triggers an additional hop at the front of
  the cluster (note that only one hop is triggered, i.e., no
  ``avalanches'' develop). (iii) A second hopping is \emph{not} triggered, however, when
  the cluster is of size $l > l_{\max}$. (b) The same model, mapped to
  a ZRP-like picture: each vacancy in the exclusion process is mapped to a
  ZRP site, and the size of the cluster to the right of the vacancy is
  mapped to the occupation of the ZRP site.}}
\end{figure}

The original definition of the AEP in Ref.\ \cite{DongAEPprl2012} is
as an exclusion process similar to the TASEP. In this model, $N$
particles are distributed among $L_\text{AEP}$ sites of a
one-dimensional lattice (we add the subscript AEP for later
convenience). The particles have an exclusion interaction which does
not allow two or more particles to reside on the same site, and
therefore each site is either empty or occupied by a single
particle. The dynamics of the model, illustrated in \fig
\ref{fig:aepscheme}a, proceeds as follows: a particle at site $i$
may hop to $i+1$ with rate (i.e., probability per unit time) 1,
provided site $i+1$ is empty (here $i=1,\ldots,L_\text{AEP}$ is an
arbitrary site). If an advancing particle joins an existing cluster
of $k$ particles, i.e. if sites $i+2,\ldots,i+k+1$ are occupied but
$i+k+2$ is empty, it may facilitate a (simultaneous) hop of the
particle at the other end of the cluster from site $i+k+1$ to
$i+k+2$. The second, facilitated, hop occurs if and only if the
cluster is of length $1 \leq k \leq \lmax$, where $\lmax \leq L$ is
a parameter of the model. The model thus has three parameters
($L_\text{AEP},N,\lmax$). In the current work we focus on the
thermodynamic limit in which $L_\text{AEP},N\to\infty$ while their
ratio, the particle density $\rho_\text{AEP}\equiv N/L_\text{AEP}$,
remains fixed. The particle current of this model is always larger
than the current of the corresponding TASEP (which is the same model
with no facilitated hops), and for this reason it was called an
accelerated exclusion process in Ref.\ \cite{DongAEPprl2012}.

We now describe the model in the ZRP picture. In this picture, there
are $L_\text{ZRP}$ sites, each occupied by a fluctuating number of
particles $n_i$ where the index $i=1,\ldots,L_\text{ZRP}$ is again
used to denote sites in the lattice. Unlike the TASEP picture, in
the ZRP picture there is no restriction on the number of particles
per site, and therefore $n_i \geq 0$ can attain any positive integer
value. The dynamics proceeds with particles hopping stochastically
among sites as follows. An occupied site $i$ ejects a particle with
rate 1. The particle then hops in a totally asymmetric fashion: if
site $i+1$ is empty, or if it has more than $\ell_\text{max}$
particles, the particle moves to $i+1$. Otherwise, i.e. if $1\leq
n_{i+1} \leq \ell_\text{max}$, the particle advances another site
and lands at site $i+2$. The dynamics can be summarized as
\begin{widetext}
\begin{align}\label{eq:AEPdynamics}
\ldots,n_i, n_{i+1},n_{i+2},\ldots &\xrightarrow{\mathbbm{1}(n_i\geq
1)} \ldots,n_i-1, n_{i+1}+1,n_{i+2},\ldots \quad \text{if }
n_{i+1}=0
\text{ or } n_{i+1} > \ell_\text{max} \nonumber \\
\ldots,n_i, n_{i+1},n_{i+2},\ldots &\xrightarrow{\mathbbm{1}(n_i\geq
1)} \ldots,n_i-1, n_{i+1},n_{i+2}+1,\ldots \quad \text{if } 1\leq
n_{i+1} \leq \ell_\text{max},
\end{align}
\end{widetext}
(see also \fig \ref{fig:aepscheme}b) where
$\mathbbm{1}(\text{condition})$ equals one if the condition is met
and zero otherwise. This dynamics conserves the total number of
particles, $N \equiv \sum_i n_i$. Once again, the model has three
parameters, which in the thermodynamic limit of $L_\text{ZRP},N \to
\infty$ with a constant ratio are reduced to the density
$\rho_\text{ZRP} \equiv N/L_\text{ZRP}$ and $\lmax$.

The ZRP picture is obtained from the AEP by focusing on the dynamics
of clusters rather than the occupation of sites. Each configuration
of an AEP with $L_\text{AEP}$ site, $N$ particles, and $H\equiv
L_\text{AEP}-N$ holes (i.e., unoccupied sites) can be mapped to a
ZRP configuration with $L_\text{ZRP}=H$ sites and $N$ particles as
follows: each hole in the AEP is mapped to a site in the ZRP, and
the size of the cluster to the right of the hole (i.e., the number
of particles between this hole and the next hole) is mapped to the
occupation of the corresponding ZRP site. This mapping is
illustrated in \fig \ref{fig:aepscheme}. It is straightforward to
verify that under this mapping, the dynamics of the AEP is exactly
mapped to the ZRP picture dynamics defined above.

Since the number of sites in the AEP and the corresponding ZRP are
not the same, the density of particles in the two also differ. The
density in one picture can be translated to that of the other
according to the relations
\begin{equation}\label{eq:TranslationRho}
\rho_\text{ZRP} = \frac{\rho_\text{AEP}}{(1-\rho_\text{AEP})} \qquad
\text{and}\qquad \rho_\text{AEP} =
\frac{\rho_\text{ZRP}}{(1+\rho_\text{ZRP})}.
\end{equation}
Similar equations relate the particle currents $J_\text{ZRP}$ and
$J_\text{AEP}$ in the two systems. In both systems, the current is
defined as the total number of sites traversed by hopping particles
per unit time divided by the respective system size. Therefore,
\begin{equation}\label{eq:TranslationJ}
J_\text{ZRP} = (1+\rho_\text{ZRP})J_\text{AEP} \quad \text{and}
\quad J_\text{AEP} = (1-\rho_\text{AEP})J_\text{ZRP}.
\end{equation}

Working in the ZRP picture is more convenient for the purpose of
studying condensation in the model. Therefore, in the rest of the
paper we concentrate on the ZRP picture and only occasionally
translate the results to the exclusion process picture. To simplify
notation, in what follows we drop the subscript ZRP and denote $L
\equiv L_\text{ZRP}$,  $\rho \equiv \rho_\text{ZRP}$ and $J \equiv
J_\text{ZRP}$.

\section{Summary of main results}\label{sec:MainResults}
The AEP was studied numerically in Ref.\ \cite{DongAEPprl2012},
where several interesting phenomena were found. The most striking
result is that for a finite system of $L_\text{AEP} = 1000$ and
several values of $\lmax$ that range from 10 to 500, there is an
apparent transition from a disordered phase at low densities to a
condensed phase at high densities. The disordered phase is
homogeneous, while in the condensed phase there is a cluster of
particles of macroscopic size, consisting of a significant fraction
of all particles. The transition is also visible in the
current-density relation: in the disordered phase, the current
changes nonlinearly with the density, while the condensed phase is
characterized by a current $J_\text{AEP} = 1-\rho_\text{AEP}$, which
corresponds to $J_\text{ZRP} = 1$.

Recently, a combination of a MF analysis, exact results, and
numerics, was used in Ref.\ \cite{DongAEPpre2013} to clarify some of
the earlier numerical findings. In particular, the model was
analyzed in the homogeneous phase, and also in the special limit of
$N \to \infty$ with $L_\text{ZRP}$ fixed (in the TASEP picture this
corresponds to a fixed number of holes). However, several questions
that arise naturally from the numerical results remain unanswered.
Is there a true phase transition in the thermodynamic limit of the
model? What is the order of the phase transition? Can one calculate
the phase diagram and explain the simple form of the current in the
condensed phase? In the present paper, we seek solutions to these
questions within a MF approximation. The results of our analysis are
as follows.

\begin{enumerate}
\item The mean-field approximation suggests that a true
    condensation phase transition occurs in the thermodynamic
    limit when $\lmax$ diverges with $L$ at least
    logarithmically, i.e., $\lmax > C \log L$, for large enough
    constant $C$. If $\lmax$ is kept constant in the limit of
    $L,N \to \infty$, no phase transition occurs.

\item The transition is continuous when $C \log L < \lmax \ll
    L$. The critical density (in the ZRP picture) is $\rho_c =
    1$, in agreement with the numerics of Ref.\
    \cite{DongAEPprl2012}. The current in the critical phase (in
    the ZRP picture) is $J_c = 1$. As in the ZRP, this current
    is independent of the density. This explains why holes have
    a unit velocity in the condensed phase, as found in Ref.\
    \cite{DongAEPprl2012}.

\item The transition becomes discontinuous when $\lmax$ scales
    linearly with $L$, i.e, when $\lmax = a L$ where $a$ is a
    constant. The density at the transition point can be
    obtained within the mean-field approximation by calculating
    the large deviation function for the size of the condensate.
    The transition density satisfies $\rho_\text{trans}(a) > 1$.
    The mean-field picture suggests that close to the transition
    there are metastability and hysteresis effects, which might
    explain the large fluctuations measured numerically in Ref.\
    \cite{DongAEPprl2012}.

\item In the condensed phase, the condensate drifts by skipping
    every other site (i.e., it drifts from site 1 to 3 to 5,
    etc.). This drift is similar to that found recently in Ref.\
    \cite{CondDriftPRE2013}. The drift velocity decays
    algebraically or faster with the system size (depending on
    the scaling of $\lmax$ with $L$), and the condensate is
    typically supported on a single site.

\item Although the MF description is usually expected to be only
    qualitatively correct, some of our MF predictions, including
    the phase diagram, quantitatively match numerical results to
    a high accuracy. It is not yet known whether the mean-field
    approximation indeed yields the exact phase diagram of the
    model, and if so, why.
\end{enumerate}

\section{Condensation transition in the AEP}\label{sec:Calculation}

In this section, we use the mean-field (MF) approximation of Ref.\
\cite{CondDriftPRE2013} to analyze the AEP condensation transition.
In MF analyses of models such as the AEP, one usually assumes that
the occupations of different sites are independent and identically
distributed. In the MF picture we employ here, the occupations of
different sites are assumed to be independent in the steady-state,
but not necessarily identically distributed. In particular, in the
condensed phase, where the condensate spontaneously breaks the
translational symmetry of the model, the occupation probability of a
site is allowed to depend on its distance from the condensate.
Indeed, due to spatial correlations, the \emph{true} marginal
occupation probability of a site is expected to vary with the
distance from the condensate. Thus, the non-homogeneous MF ansatz
which we consider effectively captures some of the effects of
correlations between sites, even though sites are ultimately assumed
to be independent. Allowing the steady state distribution to be
inhomogeneous (even though the dynamics is translationally
invariant) is the main technical novelty of our approach.

Mathematically stated, the MF assumption postulates that the
stationary distribution has a product form
\begin{equation}\label{eq:AEPMFproductGC}
P_{\rho}^{gc}(n_1,\ldots,n_L) = \prod_{i=1}^L P_{i}(n_i|\rho),
\end{equation}
where $P_i(n|\rho)$ denotes the probability that, within this
mean-field approach, site $i$ has exactly $n$ particles in the
steady state. The single site distributions $P_i(n|\rho)$ depend on
the density $\rho$ via the requirement that the mean total number of
particles satisfies
\begin{equation}\label{eq:GCmean}
\sum_{i=1}^L \sum_{n_i=0}^\infty n_i P_i(n_i|\rho) = L \rho.
\end{equation}

The ansatz (\ref{eq:AEPMFproductGC}) is ``grand-canonical'' in
nature, in the sense that the total number of particles in the
system is allowed to fluctuate around its mean (\ref{eq:GCmean}).
Such an ansatz is expected to be useful for the description of
typical fluctuations in the steady state (and possibly also in
metastable states). However, one cannot hope that it successfully
describes, for example, the probability 
that a single site accommodates all $N$ particles. More generally,
the ansatz must be adjusted in any event where the occupations of
some sites are so large that, due to the constraint on the total
particle number, a macroscopic fraction of all other sites is forced
to be in an atypical state. In order to describe such situations,
one should consider a ``canonical'' distribution, where the total
number of particles is constrained. To simplify the calculation, we
instead propose a ``hybrid'' ansatz, in which sites are split into
two groups --- macroscopically occupied sites (the condensate), and
the rest (the fluid background). The distribution of the fluid is
still taken as to be factorized, but its mean occupation depends on
the occupation of the condensate sites. For a configuration with $k$
condensate sites with occupations $n_{j_i} = m_i L$ (for
$i=1,\ldots,k$), this ansatz may be written as
\begin{equation}\label{eq:AEPMFproduct}
P_{\rho}^{c}(n_1,\ldots,n_L) = P_\text{max}^{(k)}\,({\mathbf m})
\prod_{i} P_{i}\Bigl(n_i\Big|\rho-\sum m_i\Bigr),
\end{equation}
where $P_\text{max}^{(k)}(\bf m)$ is the probability that the $k$
most occupied sites have occupations ${\mathbf m}L \equiv (m_1
L,\ldots,m_k L)$, and the product runs over all sites other than
$j_1,\ldots,j_k$. This ansatz is not adequate for the description of
the most general situations, but it shall suffice for the
calculation of the AEP phase diagram. Abusing the terminology
somewhat, we call this ansatz below the canonical ansatz. Note that
$P_\text{max}^{(k)}$ may depend not only on the occupations of the
highly occupied sites, but also on their locations. We suppress this
dependence in the notation to avoid clutter. In addition, in what
follows, the dependence of $P_i$ on $\rho$ shall also be suppressed,
i.e., we shall denote $P_i(n_i)\equiv P_i(n_i|\rho)$.

For the calculation of the phase diagram, the grand-canonical ansatz
(\ref{eq:AEPMFproductGC}) suffices if a condensate can be created
without macroscopically affecting all other sites. This happens when
$\lmax \ll L$, as the occupation of any site may in this case reach
$\lmax$ without noticeably affecting the occupations of other sites.
However, when $\lmax = a L$ (where $a$ is a finite constant), a
finite fraction of all particles in the system must be located in a
single site if its occupation is to reach $\lmax$, and the state of
\emph{all} other sites is influenced. Therefore, the canonical
ansatz (\ref{eq:AEPMFproduct}) must be employed to analyze this
case. We show below that in the first case ($\lmax \ll L$), the
condensation transition is continuous, while it becomes
discontinuous in the second case ($\lmax = a L$). The inadequacy of
the grand-canonical ansatz in the latter case is a manifestation of
an inequivalence of the two ensembles. This is similar to the
behavior of equilibrium systems with long-range interactions, where
first order phase transitions may lead to inequivalence of ensembles
\cite{LongRangeLesHouches,LongRangeJStatMech}. A condensation
transition with very a similar phenomenology, and indeed a similar
mathematical description, has been studied in Refs.\
\cite{GrosskinskySchutzZRP1stOrder,ChlebounThesis,%
ChlebounGrosskinsky2014CondensationFactorized,ChlebounGrosskinsky2014DynamicalMetastability}
in a ZRP with with rates that depend on the system size $L$. The
methods of analysis presented in these works will prove useful in
what follows.

We begin the calculation by considering, in \sect
\ref{sec:SecondOrder}, the case of $\lmax \ll L$, where the phase
diagram can be calculated using the grand-canonical ansatz
(\ref{eq:AEPMFproductGC}). The grand-canonical calculation is
simpler than the canonical one, as sites are completely independent
of each other according to the assumption (\ref{eq:AEPMFproductGC}).
Therefore, the description of the MF procedure is more transparent
in the grand-canonical ensemble. We then move on to the case of
$\lmax = a L$, where ensembles are inequivalent. The canonical
calculation in this case is presented in \sect \ref{sec:Canonical}.

\subsection{Grand-canonical calculation, $\lmax \ll L$}\label{sec:SecondOrder}

Assume that the steady-state distribution has the form
(\ref{eq:AEPMFproductGC}). Our goal is to find the single-site
marginals $P_i(n)$. Before writing down the master equation that
these marginals satisfy, we define two auxiliary quantities that
simplify subsequent notation: by $\q_i$ we denote the mean influx of
particles into site $i$ conditioned on its occupation satisfying
$1\leq n_i \leq \lmax$. Similarly, we write $\Q_i$ to denote the
mean influx conditioned on $n_i = 0$ or $n_i > \lmax$. Within the MF
approximation, these are equal to
\begin{align}\label{eq:AEPqDefinition}
\q_i &= \Bigl(\sum_{k=1}^{\lmax}
P_{i-1}(k)\Bigr)\bigl(1-P_{i-2}(0)\bigr) \nonumber \\
\Q_i &= \q_i + \bigl(1-P_{i-1}(0)\bigr) \geq \q_i.
\end{align}
The first line states that when $1\leq n_i \leq \lmax$, the only
particles that are added to site $i$ are those that leave site $i-2$
(these hop with rate 1 whenever site $i-2$ is not empty), and do not
stay at site $i-1$ (i.e., $1\leq n_{i-1} \leq \lmax$). The second
line states that when $n_i=0$ or $n_i > \lmax$, the incoming current
into site $i$ is higher because it also includes all particles which
depart from of site $i-1$.

Using this notation, the master equation reads
\begin{align}\label{eq:AEPmasterEq}
\dot{P}_i(0) ={}& P_i(1)-P_i(0)\Q_i \nonumber \\
\dot{P}_i(1) ={}& P_i(2) + P_i(0)\Q_i-P_i(1)[1+\q_i] \nonumber \\
\dot{P}_i(n) ={}& P_i(n+1) + P_i(n-1)\mathcal{Q}_i(n-1) \nonumber \\
&{} - P_i(n)[1+ \mathcal{Q}_i(n)],
\end{align}
where
\begin{equation}
\mathcal{Q}_i(n) \equiv
\begin{cases}
\q_i & \text{if } 1\leq n \leq \lmax \\
\Q_i & \text{if } n = 0 \text{ or } n \geq \lmax + 1
\end{cases}.
\end{equation}
In the steady state the left hand sides of \eqns
(\ref{eq:AEPmasterEq}) vanish and they can be recursively solved,
yielding
\begin{equation}\label{eq:AEPstationaryDist}
P_i(n) = P_i(0) \begin{cases} \Q_i \q_i^{n-1} &\text{if }
1\leq
n \leq \lmax + 1 \\
\Q_i^{n-\lmax} \q_i^{\lmax} &\text{if } n \geq \lmax + 1
\end{cases}.
\end{equation}
Assuming that $\Q_i < 1$ (an assumption that will be examined
below), the normalization condition $\sum_n P(n) = 1$ yields
\begin{equation}\label{eq:AEPp0}
P_i(0) = \frac{(1-\q_i)(1-\Q_i)}{1 - \q_i -
\Q_i(\Q_i-\q_i)(1 - \q_i^\lmax)} \simeq
\frac{1-\q_i}{1+\Q_i - \q_i},
\end{equation}
where the last approximate equality becomes exact in the limit
$\lmax \to \infty$. Similarly, the mean occupation of site $i$ is,
when $\lmax \to \infty$,
\begin{equation}\label{eq:AEPrho}
\rho_i \equiv \sum_{n} n P_i(n) \simeq \frac{\Q_i} {(1-\q_i)(1+\Q_i-\q_i)}.
\end{equation}

The distribution (\ref{eq:AEPstationaryDist}) is normalizable only
as long as $\q_i$ and $\Q_i$ are both less than 1. If $\Q_i$ (the
larger of the two) is found to be 1 or more, the above analysis is
inconsistent. Physically, such an inconsistency means that site $i$
tends to accumulate an ever-increasing number of particles,
signalling that a condensate forms on site $i$. If $\Q_i < 1$ for
all densities $\rho$, then there is no condensation transition in
the model. Below we show that this is not the case, and that a
condensation transition occurs at a critical density $\rho_c$ at
which $Q(\rho_c) = 1$. We proceed by analyzing separately the
subcritical phase, in which the system is homogeneous, and the
supercritical phase in which the condensate breaks translational
invariance.

\subsubsection{The homogeneous phase}

In the homogeneous phase, $P_i, \q_i$ and $\Q_i$ are site
independent, and therefore the subscript $i$ may be dropped. From
equations (\ref{eq:AEPqDefinition}) and (\ref{eq:AEPp0}) it is found
that when ${\lmax \to \infty}$
\begin{equation}
\Q = \q + \sqrt{\q} \quad \text{or equivalently} \quad \q =
\frac{1}{2}\bigl(1+2\Q - \sqrt{1+4\Q}\bigr).
\end{equation}
Substituting this relation in \eqn (\ref{eq:AEPrho}) yields
\begin{equation}\label{eq:HomogeneousQ}
\Q =
\frac{1-\rho+2\rho^2-(1-\rho)\sqrt{1+4\rho^2}}{2\rho^2}.
\end{equation}
It is seen that $\Q(\rho \to 1) = 1$, implying that the calculation
breaks down at a critical density
\begin{equation}\label{eq:HomogeneousRhoC}
\rho_c \equiv 1.
\end{equation}
This breakdown signals the occurrence of a condensation transition
at $\rho_c$ (remember that this is true only when the
grand-canonical ensemble is equivalent to the canonical ensemble,
i.e., when ${1 \ll \lmax \ll L}$. As discussed below, when $\lmax$
scales linearly with $L$ the condensation transition occurs at a
density $\rho_\text{trans} > \rho_c$). In the exclusion process
picture, the corresponding critical density is $\rho_{\text{AEP},c}
= 1/2$ [see \eqn (\ref{eq:TranslationRho})], which is in agreement
with the findings of Ref.\ \cite{DongAEPprl2012}.

\subsubsection{The supercritical phase}\label{sec:SecondOrderSupercritical}

The mean-field analysis of the supercritical phase begins by
assuming that there is a ``supercritical site'' whose occupation is
macroscopic. The following analysis is thus valid as long as this
site remains macroscopically occupied. On very large timescales
(which will be determined below), the condensate migrates to other
sites. Our analysis relies on the wide separation between the
timescale of the microscopic dynamics and the macroscopic timescale
of condensate motion.

Assume that a condensate is located on site 1, i.e., $n_1 \approx
\infty$. This information can be used in \eqn
(\ref{eq:AEPqDefinition}) to find that $\q_2 = 0$ and $\Q_2 = 1$,
and thus from (\ref{eq:AEPstationaryDist}) and (\ref{eq:AEPp0}) we
obtain $P_2(0) = P_2(1) = 1/2$, and $P_2(n \geq 2) = 0$. We may now
repeat the procedure to calculate $P_3(n)$ [and iteratively $P_i(n)$
for any $i$]. We thus find $\q_3 = 1/2$ and $\Q_3 = 1$. This value
of $\Q$ implies that $P_3(n)$ is not normalizable, i.e., site 3 is
``critical': once $n_3$ exceeds the value $\lmax$, the incoming
current into this site exactly equals the outgoing current, and
$n_3$ might increase and eventually take over the condensate.
However, in the condensed phase the time it takes $n_3$ to reach the
value $\lmax$ is very long --- this is in fact a requirement for the
condensed phase to exist, as we discuss below. Therefore, for long
periods of time the system is in a metastable state in which $n_3$
fluctuates around 0. We analyze the two cases $n_3 = O(1)$ and $n_3
> \lmax$ separately.

We begin by examining the long-lived metastable state in which $n_3$
remains finite. In this metastable state, $n_3$ performs a random
walk biased towards $n_3 = 0$ with a reflecting boundary condition
at the origin. This random walk has an absorbing wall at $n_3 =
\lmax + 1$. Conditioned on the walk not reaching this absorbing
wall, we find from \eqns (\ref{eq:AEPstationaryDist}) and
(\ref{eq:AEPp0}) that (for $\lmax \gg 1$) $P_3(0) \simeq 1/3$ and
$P_3(n) \simeq 2^{-(n-1)}/3$ for $n \geq 1$. Moving to site 4 we
find $\q_4 = 1/3$ and $\Q_4 = 1$, i.e., site 4 is also critical.
Once again, there is a long-lived metastable state in which $n_4$ is
finite, and eventually, when $n_4$ reaches $\lmax$, it might
increase until it takes over and becomes the new condensate. This
picture continues at all sites, and as before we first examine the
metastable state in which all sites have finite occupations, $n_i
\leq \lmax$. Continuing the procedure iteratively, we show in
Appendix \ref{sec:AppendixFib} that  $\Q_i = 1$ for all $i\geq 2$,
$\q_i = P_{i-1}(0)$, and $P_i(0) = F_{i-1}/F_{i+1}$ where $F_i$ is
the $i$'th Fibonacci number. Therefore, $\q_i$ and $P_i(0)$ converge
exponentially with $i$ to $\q_\infty \equiv (3-\sqrt{5})/2 \approx
0.38$.

We now examine what happens once a site $2\leq i \leq L$ reaches
$n_i = \lmax + 1$. As long as $n_i$ remains larger than $\lmax + 1$,
it performs an unbiased random walk. If it reaches $\lmax + 1$ the
random walk becomes biased again towards $n_i = 0$, and the
occupation rapidly decreases to its metastable, nearly-empty state.
On the other hand, if it reaches $n_i \approx N_\text{cond}-\lmax$
(where $N_\text{cond}$ is the typical number of particles in the
condensate) the old condensate becomes depleted and site $i$ takes
over and becomes the new condensate. There is of course a
possibility that while $\lmax < n_i < N_\text{cond}-\lmax$, another
site (or sites) reaches $\lmax + 1$, leading to a situation with
three (or more) highly occupied site. If this event is quite
probable, the system typically does not have just a single
condensate but many highly occupied sites (possibly a finite density
of them), and therefore the system will not be in a truly condensed
state.

In order to ensure that typically there is only one highly occupied
site at a time, and, on rare occasions, no more than two such sites,
one must choose $\lmax$ to be large enough so that the time
$T_\lmax$ that it takes until some site reaches $\lmax$ is much
larger than the time $T_\text{takeover}$ that passes before a highly
occupied site takes over the condensate. Since $\lmax \ll
N_\text{cond} = O(L)$, the latter scales as $T_\text{takeover} =
O(N_\text{cond}^2) = O(L^2)$ (this is the well-known gambler's ruin
problem for an unbiased random walk). On the other hand, $T_\lmax$
scales as
\begin{equation}\label{eq:AEPTlmax}
T_\lmax = O\Bigl(\bigl[\sum_i \q_i^{\lmax}\bigr]^{-1}\Bigr) \approx O\Bigl(\min
\bigl(\q_3^{-\lmax}, \q_\infty^{-\lmax} / L \bigr)\Bigr).
\end{equation}
The first of the two terms on the right hand site of
(\ref{eq:AEPTlmax}) corresponds to the time it takes site 3 to reach
$\lmax$ particles (the probability of site 3 to reach $\lmax$ is
higher than that of any other site because $\q_3
> \q_i$ for all sites $i$), and the second corresponds to a distant
site (with $\q_i \approx \q_\infty$) reaching $\lmax$ (although this
probability is smaller than that of site 3, there are $O(L)$ such
sites, increasing the probability that one of them reaches $\lmax$).
The condition $T_\lmax \gg T_\text{takeover}$ then implies that
\begin{equation}
\lmax \gg \frac{3}{-\log \q_\infty}\, \log L \approx 3.12\, \log L
\end{equation}
must hold in order for the system to have a single condensate. This
means that a true condensation transition takes place only when
$\lmax$ increases logarithmically (where the logarithm has a large
enough prefactor) or faster with the system size. The same criterion
for condensation was suggested in a similar model in Ref.\
\cite{GrosskinskySchutzZRP1stOrder}

\subsubsection{Dynamics of the
condensate}\label{sec:SecondOrderDynamics}

We now discuss the dynamics of the condensate. We begin by noting
that when $\log L \lesssim \lmax \ll L$, it is highly improbable for
an AEP with a supercritical density to be (momentarily) in a
homogeneous disordered state, i.e., where no occupation exceeds
$\lmax$. This fact can once again be understood dynamically: the
condensate must lose $O(L)$ particles without a new condensate
forming in order for the system to reach such a disordered state.
The typical timescale in which this process occurs is exponentially
large: $e^{c L}$ for some constant $c$, essentially since this
process necessitates that \emph{all} other (independent) sites are
to be atypically occupied (see below in \sect \ref{sec:Canonical}).
On the other hand, if the system is in a disordered state, it takes
a time of order $O(q_i^{-\lmax})\ll e^{c L}$ for a site to reach
$\lmax$ and a condensate to form. Therefore, the fraction of time
the system spends in a disordered state is negligible in the
thermodynamic limit.

Thus, the dynamics of the condensate is dominated by events where an
additional condensate forms on another site, eventually taking over
the old one. Where and when does this new condensate form? First,
consider the case that $\lmax = A \log L$ with $A$ large enough so
that typically there is indeed a single condensate. As discussed
above, from time to time a fluctuation may cause another site to
reach $\lmax$ and (with some probability) to take over the
condensate. What is the most probable location of the new
condensate? To answer this question, compare the time it takes site
3 to reach $\lmax$, which is of order $O(L^{-A\log \q_3})$, with the
time it takes a distant site to reach $\lmax$, which is of order
$O(L^{-A \log \q_\infty -1})$ (remember that there are $O(L)$ such
distant sites). Comparing these, we find that when $A >
[\log(\q_3/\q_\infty)]^{-1} \approx 3.71$ the new condensate forms
most frequently on site 3, i.e., two sites downstream from the
current condensate. In this case, the condensate performs a drift
motion, skipping every other site. In the thermodynamic limit,
however, the drift velocity of the condensate decreases to zero
algebraically with the system size, as $L^{-A\log 2}$.

When $\log L \ll \lmax \ll L$ (e.g., $\lmax \sim \sqrt{L}$), the
argument of the previous paragraph shows that a condensate at site
$i$ always relocates to site $i+2$, and the resulting drift velocity
is of order $2^{-\lmax}$. This velocity decreases with the system
size faster than algebraically but slower than exponentially (e.g.,
as a stretched exponential when $\lmax \sim \sqrt{L}$).


\subsection{Canonical calculation, $\lmax = a L$}\label{sec:Canonical}

As discussed above, when $\lmax = a L$ the analysis must proceed in
the canonical ensemble. We now show that in this case the
condensation phase transition is of first, rather than second,
order. Thus, the canonical and grand-canonical ensembles are
inequivalent.
Our analysis of this case follows the ideas of Ref.\
\cite{ChlebounThesis}: to find the density at which there is a phase
transition, we shall calculate the occupation probability of the
most occupied site, or more precisely, the large-deviation function
(LDF) for this occupation \footnote{If $P(n_{\max} = m L) \sim e^{-L
I(m)}$ then $I(m)$ is called the large-deviation function or rate
function of $n_{\max}$; see \cite{TouchetteReview} for an
introduction to the theory of large deviations. Throughout the paper
we use the symbol $\sim$ to denote exponential equivalence to
leading order in $L$. For example, the above relation means that
$I(m) = \lim_{L\to\infty}L^{-1}\log P({n_{\max} = m L})$.}. This LDF
is found by examining the dynamics of the condensate occupation. The
condensate LDF plays a role similar to that of an equilibrium free
energy, and thus once it is found a Landau-theory-type analysis
yields the phase diagram of the model.

\subsubsection{Calculation of the condensate
LDF}\label{sec:CondensateLDFcalculation}

The calculation of the LDF proceeds in three steps: (i) we express
the LDF in terms of the currents entering and leaving the
condensate; (ii) we express these currents as a function of the
density of the background fluid; and (iii) we discuss how the
background density depends on the occupation of the condensate.

\emph{Step (i)}. Assume that the condensate has $n_{\max} = m L$
particles (here we use the word condensate to mean the most occupied
site). Denote by $\jin(m)$ the mean momentary current flowing into
the condensate conditioned on this occupation, and similarly by
$\jout(m)$ the mean current flowing out of the condensate. Within
the canonical MF ansatz (\ref{eq:AEPMFproduct}), these currents
might depend on $m$ through the constraint on the total number of
particles, but there are no further correlations between the
condensate and the rest of the system. In other words, the rest of
the system may determine the functions $\jin(m)$ and $\jout(m)$, but
otherwise one may consider the dynamics of the condensate separately
from that of the rest of the system. Denoting $P_{\max}(n) \equiv
P(n_{\max}=n)$, the master equation for the condensate occupation is
thus
\begin{align}\label{eq:LDFmasterEq}
\dot{P}_{\max}(mL) &= P_{\max}(mL-1) \jin(m-L^{-1}) \nonumber \\
&\hphantom{{} = }{} + P_{\max}(mL+1)\jout(m+L^{-1}) \nonumber \\
&\hphantom{{} = }{} - P_{\max}(mL)\bigl[\jin(m)+\jout(m)\bigr].
\end{align}
Equating the left-hand side of (\ref{eq:LDFmasterEq}) to zero, and
substituting the LDF ansatz
\begin{equation}\label{eq:LDFansatz}
P_{\max}(mL) \sim e^{-L I_\rho(m)}
\end{equation}
yields to leading order in $L$
\begin{align}\label{eq:LDFwkbEq}
0 &\sim e^{I_{\rho}'(m)}\jin(m) + e^{-I_{\rho}'(m)}\jout(m) - [\jin(m) + \jout(m)] \nonumber \\
  &=\bigl(1-e^{I_{\rho}'(m)}\bigr)\,\bigl(e^{-I_{\rho}'(m)}\jout(m)-\jin(m)\bigr)
\end{align}
(this is similar to the WKB approximation in quantum mechanics).
Since the first brackets on the right-hand side cannot be
identically zero, we obtain
\begin{equation}\label{eq:LDFintegral}
I_{\rho}(m) = -\int^m \log \frac{\jin(m')}{\jout(m')}\, dm' + C,
\end{equation}
where $C$ is an integration constant which can be obtained by the
normalization requirement $\min I_{\rho}(m) = 0$.

\emph{Step (ii)}. The goal now is to find the functions $\jin(m)$
and $\jout(m)$. The latter is simply $\jout(m) = 1$, as the
departure of particles from the condensate is independent of its
size and of the rest of the system. To find the former, we assume
that at any value of condensate occupation $m$, the background fluid
can be described by \eqns
(\ref{eq:AEPstationaryDist})--(\ref{eq:AEPrho}). This would be the
case, for instance, if the relaxation timescale of the background
fluid is much shorter than the timescale in which the condensate
density $m=n_{\max}/L$ changes. Although we cannot justify this
assumption a priori, we will see below that predictions of the
ensuing calculation are very close to results measured in numerical
simulations.

We continue by obtaining from \eqns (\ref{eq:AEPqDefinition}) and
(\ref{eq:AEPp0}) a recursion relation for $P_i(0)$:
\begin{equation}\label{eq:AEPp0Recursion}
P_{i+1}(0) = \frac{P_i(0) + P_{i-1}(0) - P_i(0)P_{i-1}(0)}{2- P_i(0)}.
\end{equation}
This recursion relation is analyzed in Appendix
\ref{sec:AppendixFixedPoint}, where it is shown that all values
$0<P_\infty(0)<1$ are fixed points of this map, and furthermore the
map is exponentially contracting towards these fixed points. Thus,
the bulk of the fluid background is effectively described by the
appropriate fixed point, with deviations only in a finite boundary
layer near the condensate. The fixed point is dictated by the
condition $\rho_\infty = \rho_\text{bg}(m)$, where [from \eqns
(\ref{eq:AEPqDefinition}), (\ref{eq:AEPp0}) and (\ref{eq:AEPrho})]
\begin{equation}
P_\infty(0) = \frac{1+2\rho_\infty - \sqrt{1+4 \rho_\infty^2}}{2\rho_\infty}.
\end{equation}
The density of particles in the background fluid,
$\rho_\text{bg}(m)$, will be discussed in step (iii).

The condensate is located in site $1$, which is also site $L+1$ (due
to the periodic boundary conditions). Therefore, the current into
the condensate is $q_{L+1} \simeq q_\infty$ when ${n_{\max}<\lmax}$,
i.e., when $m<a$, and it is $Q_{L+1} \simeq Q_\infty$ when $m>a$.
Using \eqn (\ref{eq:AEPqDefinition}), we finally arrive at
\begin{equation}\label{eq:LDFjin}
\jin(m) =
1-\frac{S-1}{2\rhobg^2}\times\begin{cases}
1 & \text{if } m<a \\
1-\rhobg & \text{if } m>a
\end{cases},
\end{equation}
where $S = \sqrt{1+4\rhobg^2}$, and $\rhobg = \rhobg(m)$ is the
density of the background fluid.

\begin{figure}
        \centering
        \includegraphics[width=0.3\textwidth]{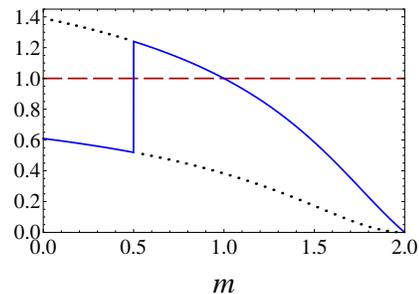}
        \caption{The incoming current into the condensate $\jin(m)$ (solid line)
        and the outflowing current $\jout(m)$ (dashed line) as a function of
        the condensate fraction $m$, for $\rho = 2$ and $a = 0.5$. The two branches
        of the function $\jin(m)$ [\eqn (\ref{eq:LDFjin})] are indicated by dotted
        lines.
        \label{fig:LDFcurrents} }
\end{figure}

The functions $\jin(m)$ and $\jout(m)$ are plotted in \fig
\ref{fig:LDFcurrents} for $\rho = 2 > \rho_c$, assuming (as we
shall, see next step) that $\rhobg(m) = \rho - m$. We see that there
is one point where $\jin(m) = \jout(m)$: at $m = \rho - 1$. At this
point, the current entering the condensate equals that leaving it,
and thus the condensate occupation is fixed. Furthermore, this fixed
point is locally stable: if $m$ decreases the incoming current
increases and vice versa. Note, however, that this fixed point
exists only if $\rho - 1 > a$ [as the jump in the curve of $\jin(m)$
occurs at $m=a$]. In addition, $m=0$ is always a locally stable
fixed point, due to the boundary condition $\jout(0) = 0$
\footnote{More precisely, for any system of finite size $L$, the
maximal occupation in the homogeneous fluid background is of order
$\log L$ (as this is the maximum of $O(L)$ independent random
variables whose distribution has an exponential tail). Therefore,
$\jout(n_{\max}/L)$ decreases from $1$ to $0$ as $n_{\max}$
decreases from some value of order $\log L$ to 0. Thus, the other
locally stable fixed point is at $m = O(\log L/L) \to 0$ when $L\to
\infty$.}. Thus we see that when $\rho < 1+a$ the only stable
solution is a disordered phase with no condensate ($m=0$), while
when $\rho
> 1 + a$ there are two locally stable states: a disordered phase,
and a condensed phase where a condensate of size $m=\rho-1$ coexists
with a background of density $\rhobg = 1 = \rho_c$. To find out
which of these dominates in the thermodynamics limit one must carry
out the integration of \eqn (\ref{eq:LDFintegral}).

\emph{Step (iii)}. The remaining task is to find the function
$\rhobg(m)$. Naively, one might assume that all particles which are
not in the condensate are in the background, i.e., $\rhobg(m) = \rho
- m$. However, when the condensate occupation $m$ becomes much lower
than its typical value, the excess particles (or at least some of
them) might not accumulate in the background fluid, but rather in a
single site (or a few sites), thus forming new condensates. Denote
by $m_2\leq m$ the fraction of particles in the second most occupied
site. The occupation probability of the two most occupied sites
again has a large deviations form
\begin{equation}
P_{\max}^{(2)}(m L, m_2 L) \sim e^{-L I_{\rho}^{(2)}(m,m_2)},
\end{equation}
and the condensate LDF is given by contraction:
\begin{equation}\label{eq:LDFcontraction}
I_\rho(m) = \min_{m_2} I_{\rho}^{(2)}(m,m_2) = I_{\rho}^{(2)}\bigl(m,m_2(m)\bigr),
\end{equation}
where $m_2(m)$ is the value of $m_2$ which achieves the minimum of
$I_{\rho}^{(2)}$ for a given value of $m$. Solving this minimization
problem is a difficult task --- instead of solving the ODE
(\ref{eq:LDFwkbEq}) one must solve a PDE which is a two-dimensional
version of this equation, see Appendix \ref{sec:Appendix2dLDF}.
Furthermore, some values of $m$ and $m_2$ are most probably achieved
by other macroscopically occupied sites forming, with occupations
$m_3 L, m_4L$, etc.. To study these, one must analyze even higher
dimensional LDFs of the form $I_{\rho}^{(k)}(m,m_2,m_3,\ldots,m_k)$.
In light of this discussion, it is seen that the background density
is $\rhobg(m) = \rho - m - m_2(m) - m_3(m) - \ldots$.

Here we shall not go through this higher dimensional analysis to
compute the exact form of $I_\rho(m)$. Instead, we first substitute
\begin{equation}\label{eq:LDFrhobgApprox}
\rhobg(m) = \rho - m
\end{equation}
and compute the LDF $\tilde{I}_\rho(m)$ when the system is
constrained to have no more than one condensate site. This LDF is an
upper bound on the true, unconstrained LDF:
\begin{equation}
\tilde{I}_\rho(m) = I_\rho^{(2)}(m,0) \geq I_\rho(m)
\end{equation}
[see (\ref{eq:LDFcontraction}) and Appendix
\ref{sec:Appendix2dLDF}]. In \sect \ref{sec:MultipleCondensates}
below we shall show that for a large range of $m$ values the two are
equal, $\tilde{I}_\rho(m) = I_\rho(m)$. In particular, we argue that
$\tilde{I}_\rho(m)$ suffices for the calculation of the phase
diagram.

\begin{figure}
        \centering
        \includegraphics[width=0.4\textwidth]{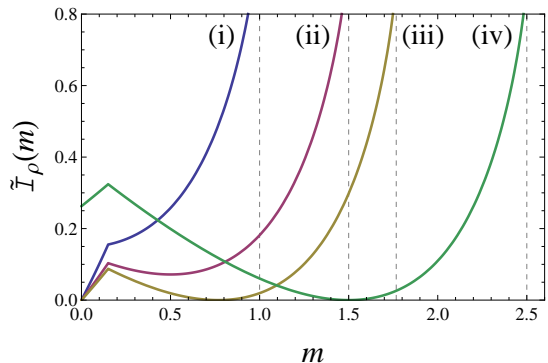}
        \caption{The large-deviation function $\tilde{I}_\rho(m)$ of the
        occupation fraction of the most occupied site for
        $a = 0.15$ and different densities:
        (i) $\rho = 1$,
        (ii) $\rho = 1.5$, (iii) $\rho = \rho_\text{trans} \approx 1.77$,
        (iv) $\rho = 2.5$. The first order transition is clearly seen:
        at densities below $\rho_\text{trans}$ the global minimum is at $m=0$, while above
        $\rho_\text{trans}$ it is at $m=\rho-1$.
        \label{fig:LDF} }
\end{figure}

Combining \eqns (\ref{eq:LDFintegral}), (\ref{eq:LDFjin}) and
(\ref{eq:LDFrhobgApprox}) yields the desired (upper bound on the)
LDF:
\begin{equation}\label{eq:LDFfinal}
\tilde{I}_\rho(m) = \rhobg(m) \log \jin(m) + A(m) + C(a,\rho),
\end{equation}
where
\begin{equation}
A(m) \equiv
\begin{cases}
\log(S-2\rhobg) + c(a,\rho)& \text{if } m<a \\
[3\log(S-2\rhobg)+1+2\rhobg-S]/4 & \text{if } m>a
\end{cases},
\end{equation}
$S = \sqrt{1+4\rhobg(m)^2}$ as before, $\rhobg \equiv \rhobg(m)$ is
given in (\ref{eq:LDFrhobgApprox}), and $c,C$ are two integration
constants. The value of $c(a,\rho)$ is chosen so that
$\tilde{I}_\rho(m)$ is continuous at ${m=a}$, while $C(a,\rho)$ is
chosen to ensure the normalization $\min_m \tilde{I}_\rho(m) = 0$
(the exact expressions for these integration constants are not
reproduced here). In \fig \ref{fig:LDF}, we plot $\tilde{I}_\rho(m)$
for a few values of $\rho$.

\subsubsection{The mean-field phase diagram}

\begin{figure}
        \centering
        \includegraphics[width=0.4\textwidth]{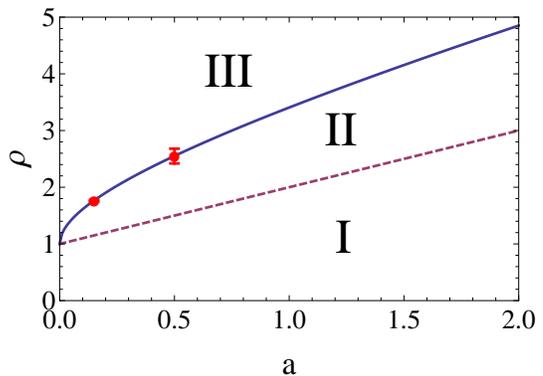}
        \caption{The mean-field phase diagram.
        Regions I and II are the disordered fluid phase, and
        region III is the condensed phase. The solid
        line marks the first order transition line $\rho_\text{trans}(a)$,
        and the dashed line, $\rho = a+1$, marks the edge of stability of
        the condensed phase. Accordingly, in region II the condensed
        phase is metastable, while it is unstable in region I.
        The two dots with error bars (at $a=0.15$ and $a=0.5$)
        were obtained from simulations
        of the model, see \sect \ref{sec:numerical} below.
        \label{fig:PhaseDiagram} }
\end{figure}

We are at last in a position to analyze the MF phase diagram of the
AEP (using the LDF $\tilde{I}_\rho(m)$; we postpone showing that it
indeed yields the correct MF phase diagram to the next paragraph).
The typical value of $m$ is the one which attains the global minimum
of the LDF (\ref{eq:LDFfinal}), which we denote by $m^*$. All other
values of $m$ are exponentially unlikely in $L$. Any local minimum
of $\tilde{I}_\rho$ other than $m^*$ is a metastable state. Such
metastable states, although unlikely, have a lifetime that grows
exponentially with $L$, as the system must overcome an exponential
barrier before the condensate occupation fraction can reach $m^*$.
The local minima are precisely those found above: studying the LDF
(\ref{eq:LDFfinal}), it is seen that the $m<a$ branch of $\tilde{I}$
is a monotonically increasing function and thus $m=0$ is always a
local minimum, while the $m>a$ branch is a convex function with a
minimum at $m= \rho-\rho_c$ [recall that $\rho_c = 1$, see
(\ref{eq:HomogeneousRhoC})]. The second local minimum exists, of
course, only if $\rho > 1+a$, otherwise this minimum falls outside
the domain $m>a$. The thermodynamic transition point
$\rho_\text{trans}$ occurs when these two minima are equal, i.e., it
is implicitly given by the equation
\begin{equation}\label{eq:LDFrhoTrans}
\tilde{I}_{\rho_\text{trans}}(0)=\tilde{I}_{\rho_\text{trans}}(\rho_\text{trans}-1).
\end{equation}
The resulting phase diagram is presented in \fig
\ref{fig:PhaseDiagram}. Note that in the limit of $a \to 0$, the
first order phase transition becomes a second order one at $\rho_c$,
as discussed above in \sect \ref{sec:SecondOrder}.

\subsubsection{The condensate LDF with multiple
condensates}\label{sec:MultipleCondensates}

We now discuss the condensate occupation LDF $I_\rho(m)$ when there
are multiple condensate sites, and argue that the upper bound
$\tilde{I}_\rho(m)$ suffices for the calculation of the phase
diagram. As explained above, $\tilde{I}_\rho(m)$ is calculated under
the assumption (\ref{eq:LDFrhobgApprox}), i.e., that there is almost
surely only one condensate site for any value of $m$. This
assumption is clearly correct when $m<a$, because then all
occupations are below $\lmax$ and the balance of incoming and
outgoing particle currents drive all sites towards the background
density (see \fig \ref{fig:LDFcurrents}). It is thus exponentially
unlikely (in $L$) to have $m_2>0$ in this regime. Similarly, \eqn
(\ref{eq:LDFrhobgApprox}) also holds when $m>\rho-1$, because in
this case the background density satisfies $\rhobg(m)\leq
\rho-m<1\equiv \rho_c$, and thus $Q_\infty(m)<1$ [see \eqn
(\ref{eq:HomogeneousQ})], i.e., the background fluid is subcritical.

\begin{figure}
        \centering
        \includegraphics[width=0.4\textwidth]{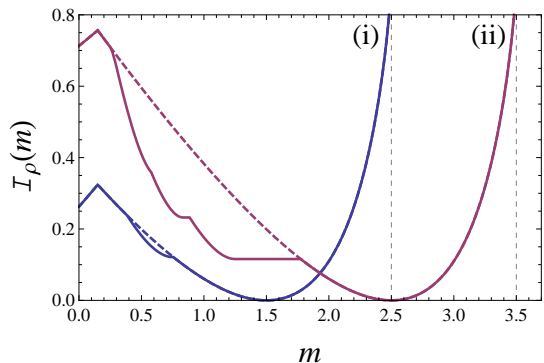}
        \caption{A schematic illustration of the LDF $I_\rho(m)$
        of the condensate occupation fraction (solid line), compared with
        the upper bound $\tilde{I}_\rho(m)$ (\ref{eq:LDFfinal}) (dashed line). The plot
        is for $a = 0.15$ and densities (i) $\rho = 2.5$, and (ii) $\rho = 3.5$. At high enough
        densities, more condensates may form as $m$ is conditioned to have increasingly
        atypical (i.e., lower) values. This is illustrated (schematically)
        by depicting a second condensate appearing
        when $\rho = 2.5$, and up to 4 condensates when $\rho = 3.5$.
        A non-analyticity of $I_\rho(m)$ is manifest at each
        value of $m$ where a new condensate appears. As explained in
        the text, $I_\rho(m) = \tilde{I}_\rho(m)$ near both local minima, and thus
        $\tilde{I}_\rho(m)$ suffices for the calculation of the phase diagram.
        \label{fig:LDFmultipleConds} }
\end{figure}

When $a<m<\rho-1$, \eqn (\ref{eq:LDFrhobgApprox}) leads to
$\rhobg(m)>1$ and thus $Q_\infty>1$. In this case one might worry
that new condensates could form on other sites. However, as
explained throughout this section, when $\lmax = a L$ the
condensation transition is first order, and the formation of a
condensate does not happen at $\rho_c$ but at a higher transition
density. Therefore, a second condensate appears only when $\rho-m >
\rho_{\text{trans},2} > 1$ for some transition density
$\rho_{\text{trans},2}$. Similarly, a third condensate site appears
only when $\rho-m-m_2 > \rho_{\text{trans},3} > 1$ (where $m_2$ is
the fraction of particles in the second condensate site), and so on.
For any value of $\rho$ and $m$, the number of condensate sites that
typically appears is finite \cite{ChlebounThesis}. A schematic
illustration of $I_\rho(m)$ at high values of $\rho$ where multiple
condensates may appear is presented in \fig
\ref{fig:LDFmultipleConds}. We conclude that $\tilde{I}_\rho(m) =
I_\rho(m)$ around both minima of $\tilde{I}_\rho(m)$ (i.e., around
$m=0$ and $m = \rho-1$), and thus the transition point calculated in
(\ref{eq:LDFrhoTrans}) is correct (within the MF approximation).

\subsubsection{Dynamics of the condensate}

When $\lmax = a L$, there are two different regimes of condensate
dynamics. At densities $\rho$ just above the transition density
$\rho_\text{trans}$, the system occasionally switches to the
disordered metastable state with $m=0$. In this regime, the
condensate dissolves, the system spends some time in a disordered
state with no condensate, and then another condensate forms on a
random site. At higher densities, the behavior is similar to that
discussed in \sect \ref{sec:SecondOrderDynamics}: a new condensate
begins to form two sites ahead of the condensate (at site 3 if the
condensate is located at site 1) while the old condensate still
exists. The probabilities of both types of events (condensate
``melting'' and condensate relocation to the next-nearest neighbor)
vanish exponentially with the system size, but with different
exponential rates which depend on $\rho$ and $a$. In the
thermodynamic limit the less improbable of the two events dominates
and dictates the typical condensate motion regime. The two dynamical
regimes are separated by a sharp dynamical phase transition, at a
density which can in principle be computed from
$I_\rho^{(2)}(m,m_2)$ \cite{ChlebounThesis}.

\subsection{Further comments about the condensed
phase}\label{sec:CondensedPhaseComments}

Two final remarks about the condensed phase. First, note that the
mean current in the condensed phase is the sum of the rate with
which a particle hops to the next site, plus twice the rate with
which it hops two sites. Therefore,
\begin{align}
J_i &= [1-P_i(0)]P_{i+1}(0) + 2[1-P_i(0)][1-P_{i+1}(0)] \nonumber \\
&= [1-P_i(0)][2-P_{i+1}(0)],
\end{align}
For all sites but a few which are close to the condensate this
current is approximately $J \simeq [1-\q_\infty][2-\q_\infty] = 1$.
This implies that the current in the corresponding AEP is
$J_\text{AEP} = 1-\rho_\text{AEP}$, as found in Ref.\
\cite{GrosskinskySchutzZRP1stOrder}.

A second remark concerns the single site occupation probability
$P(n)$, i.e., the probability the an arbitrary site has $n$
particles. According to the analysis presented above, on a finite
system of size $L$, for $n<\lmax$ this probability decays
exponentially as $P(n) \sim [\q_\infty^n + L^{-1}\q_3^n]$, while for
$n>\lmax$ it has a peak around ${N_\text{cond} = L(\rho-\rho_c)}$
due to the condensate. In between, we expect a plateau which
reflects the times in which there are two (next-nearest neighbor)
sites competing to be the condensate. The hight of this plateau
should be of order $O[\q_3^{\lmax}/(\rho-\rho_c-a)]$, where $a =
\lim_{L\to\infty}\lmax/L$. The reason for this scaling is that the
probability to see two condensate sites is approximately
$T_\text{takeover}/T_\lmax = O(L^2 \q_3^{\lmax})$, the probability
that a given site is one of these two condensates is $2/L$, and when
there are two condensate sites all $L(\rho-\rho_c-a)$ occupation
values between $\lmax$ and $N_\text{cond}-\lmax$ are equally
probable.


\section{Numerical simulations}\label{sec:numerical}
In this section, we present results of numerical simulations of the
AEP, and compare them to the MF predictions discussed above. We find
that simulation results qualitatively follow the MF predictions,
even though there are quantitative discrepancies between the two.
Furthermore, the numerical measurements of the LDF $I_\rho(m)$ and
of the phase diagram quantitatively agree, to a rather high
accuracy, with the MF theoretical predictions. This quantitative
agreement is somewhat surprising, as the MF approximation is
expected to be inexact and hold only on a qualitative level. We do
not, however, have enough numerical data to determine whether the MF
phase diagram is indeed exact.

We focus in this section on simulations of the case with a
first-order phase transition, $\lmax = a L$. The other case, of
$\log L \lesssim \lmax \ll L$ is equivalent to the limit of $a\to 0$
of the case we focus on.

\subsection{Typical behavior in the condensed phase}

\begin{figure}
        \centering
        \includegraphics[width=0.4\textwidth]{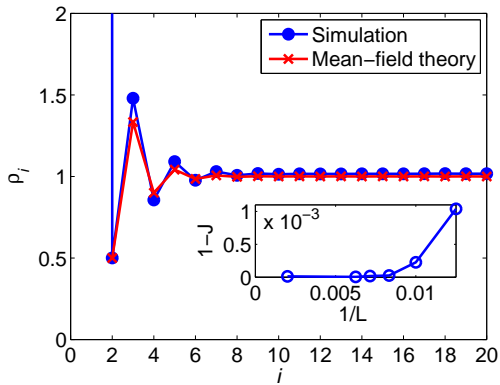}
        \caption{Numerical results and MF predictions
        for the mean density profile as seen from the condensate. The condensate
        is located at site 1 (by definition of the site labels). Simulation results are for $a = 0.15$, $\rho = 4$ and
        $L = 160$. A good qualitative agreement is observed. In the inset, the
        mean current $J$ is seen to approach $J=1$, the MF
        predicted value, when the system size is increased.
        Results are for $a=0.15$, $\rho = 4$, and various system sizes between $L=80$ and
        $L = 500$.
        \label{fig:DensityProfile} }
\end{figure}

We begin by examining the typical behavior of the AEP in the
condensed phase. First, we measure the mean density profile as seen
from the condensate: we label the position of the most occupied site
as 1, and measure $\rho_i = \langle n_i \rangle$ for $i =
1,\ldots,L$. Here $\langle \cdot \rangle$ denotes an average in the
steady state. In \fig \ref{fig:DensityProfile}, numerical results
are compared with the MF values calculated in \sect
\ref{sec:SecondOrderSupercritical}. A good qualitative agreement is
found. In particular, the density profile is seen to depend on the
distance from the condensate --- the basic fact which underlies the
condensate drift --- and furthermore the density oscillates and
decays exponentially with the distance to a value $\rho_\infty
\approx 1$. As expected, there are quantitative discrepancies
between the MF approximation and the simulation results. Note that
we have measured a small deviation $\rho_\infty \simeq 1.017$
(obtained for $a=0.15$ and $\rho=4$) from the MF prediction
$\rho_\infty = \rho_c = 1$, a deviation smaller than 2\%. This
deviation does not seem to vanish when increasing the system size up
to $L=500$ (the largest system examined).

In the inset of \fig \ref{fig:DensityProfile} we display the mean
current $J \equiv L^{-1} \sum J_i$ (where $J_i$ is the mean current
leaving site $i$) and show that it approaches the predicted value $J
= [1-\q_\infty][2-\q_\infty] = 1$.

\begin{figure}
        \centering
        \includegraphics[width=0.4\textwidth]{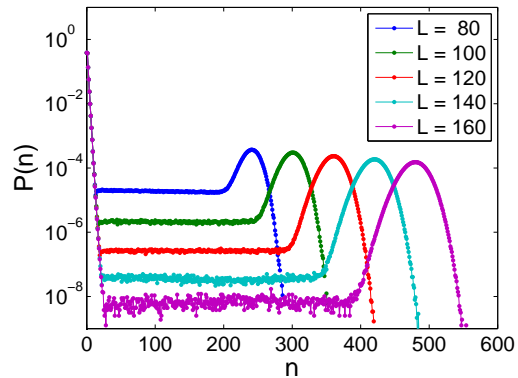}
        \caption{(Color Online) The single-site occupation probability $P(n)$ in the condensed
        phase as measured in simulations. $L$ increases from
        top to bottom. The peak due to the condensate is
        clearly seen. The plateau at intermediate values of $n$ is a consequence
        of events in which the condensate relocates to a different site, as
        discussed in \sect \ref{sec:CondensedPhaseComments}.
        Here $a = 0.15$, $\rho = 4$.
        \label{fig:OccupationProbability} }
\end{figure}

We also plot, in \fig \ref{fig:OccupationProbability}, the
single-site occupation probability $P(n)$, i.e., the probability to
find $n$ particles in any given site, in the condensed phase. As
discussed in \sect \ref{sec:CondensedPhaseComments}, $P(n)$ is
exponentially decaying for small values of $n$ --- this is the
occupation probability in the fluid background --- and has a peak at
high values of $n$ due to the condensate. The hight of the plateau
in intermediate values of $n$, which results from condensate
relocation events, decays exponentially with the system size, albeit
with a different exponent than the MF prediction discussed above
(numerical results not reproduced here).

\subsection{Phase diagram and condensate LDF}

We move on to examine the phase diagram. Via a finite-size scaling
analysis, we have directly measured $\rho_\text{trans}(a)$ for
$a=0.15$ and $a=0.5$. The numerical simulations required for this
measurement are quite lengthy, as statistics from many switches
between the $m=0$ and $m=\rho-1$ metastable states are needed, and
the typical time between such switches diverges exponentially both
with $L$ and with $a$. The details of the numerical analysis are
presented in Appendix \ref{sec:AppendixNumericalPhaseDiagram}. The
results are presented in \fig \ref{fig:PhaseDiagram}, where a very
good agreement is found between the measurements and the MF values.

\begin{figure*}
        \centering
        \includegraphics[width=0.32\textwidth]{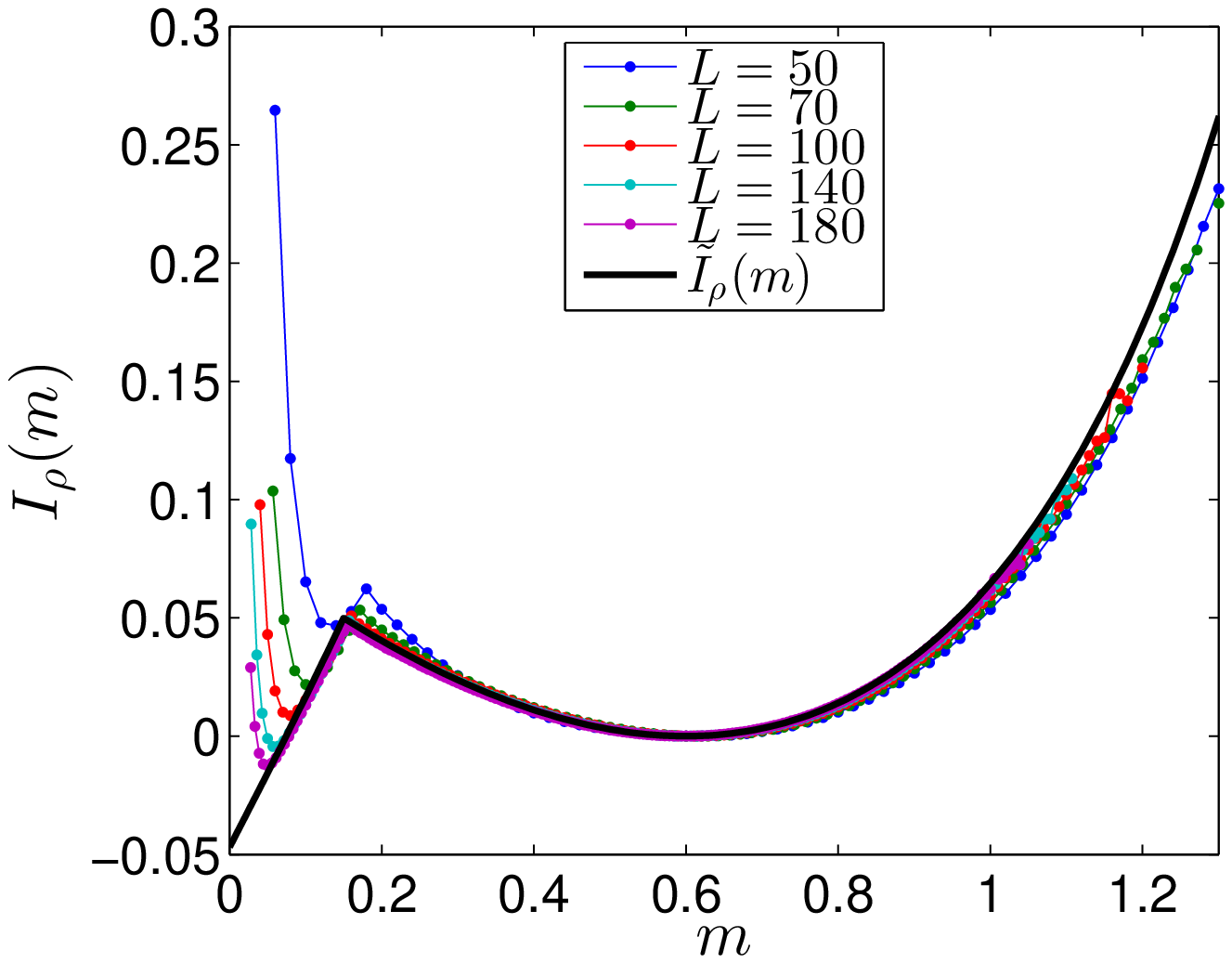}
        \includegraphics[width=0.32\textwidth]{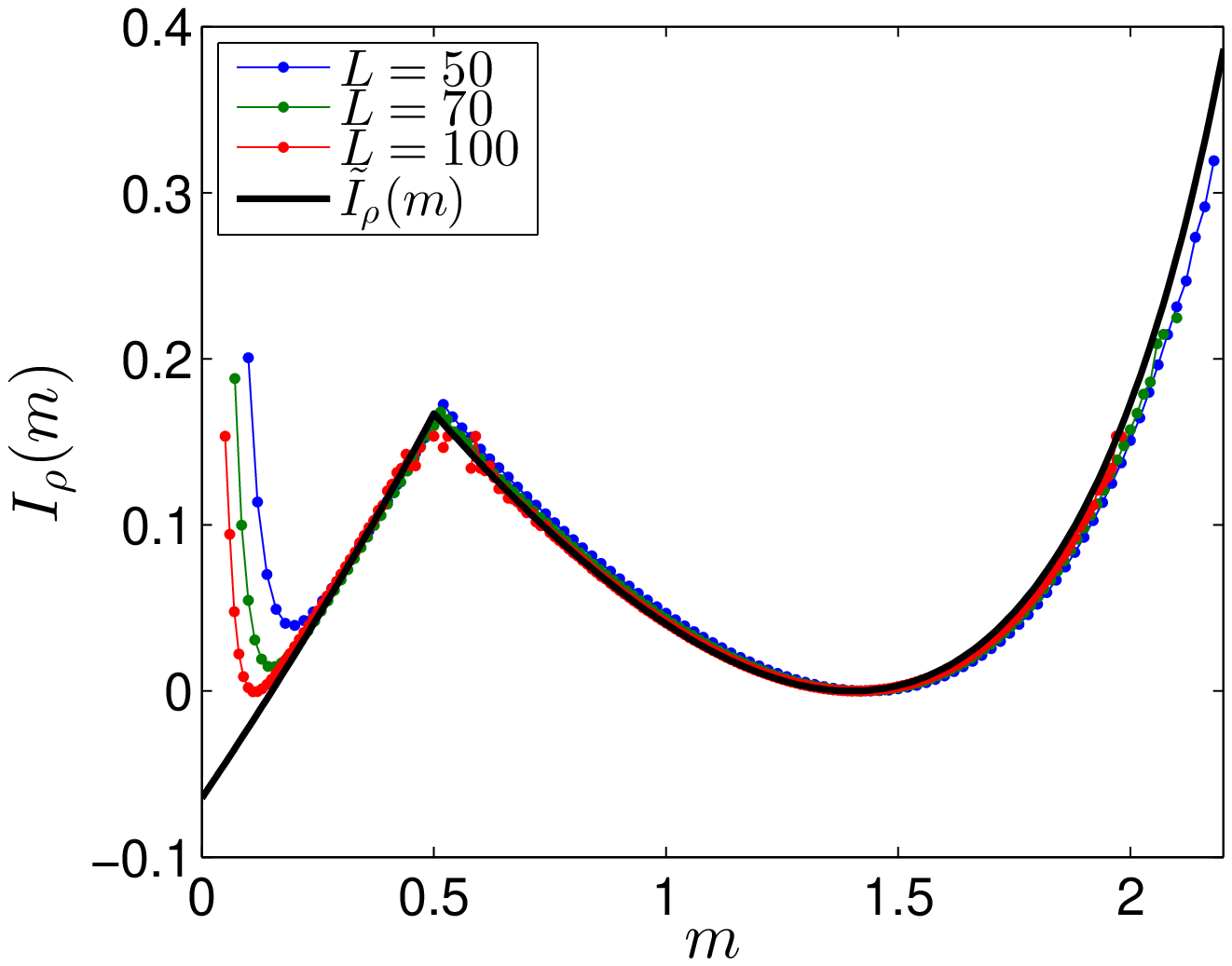}
        \includegraphics[width=0.32\textwidth]{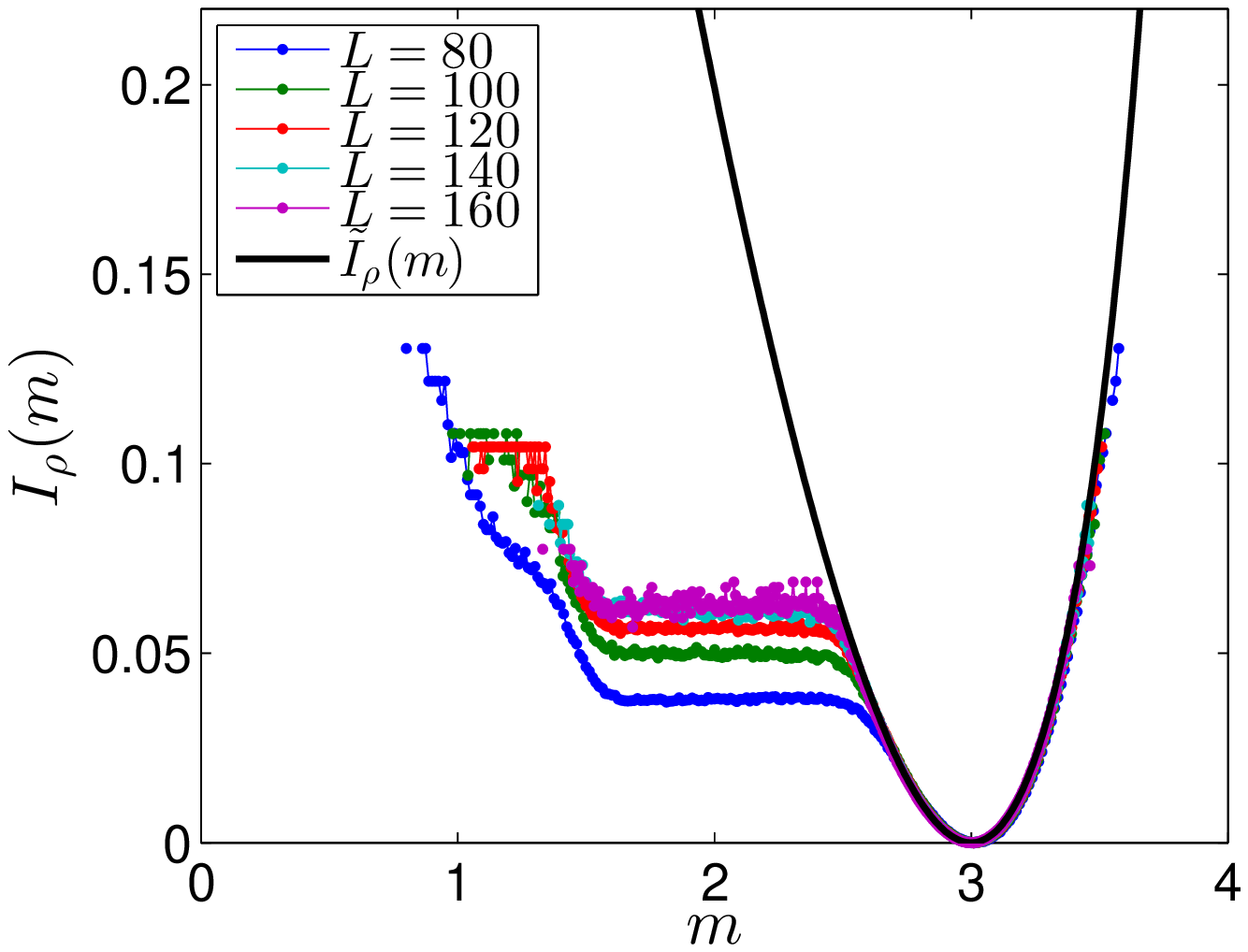}
        \caption{(Color online) Symbols: the finite-size LDF
        $I_\rho(m;L)$ [\eqn (\ref{eq:LDFfiniteDef}); lines are presented
        as a guide to the eye], measured numerically at different
        system sizes. Solid black line: the MF prediction for $\tilde{I}_\rho(m)$
        (\ref{eq:LDFfinal}) for
        $L=\infty$. All plots are normalized so that the minimum of
        the $m>a$ branch lies at $I_\rho(m) = 0$. The system size
        $L$ increases from top to bottom in the left and middle panels and
        from bottom to top in the right panel. Model parameters are (left) $a=0.15$, $\rho = 1.6$;
        (middle) $a=0.5$, $\rho = 2.4$; and (right) $a=0.15$, $\rho =
        4$. A good agreement is found between the MF predictions and
        simulation results. At high densities (the right
        plot), deviations from $\tilde{I}_\rho(m)$ due to the
        formation of multiple condensates are seen (compare with \fig
        \ref{fig:LDFmultipleConds}).
        \label{fig:NumericalLDF} }
\end{figure*}

The possibility that the MF phase diagram is exact receives stronger
support by examining the condensate occupation LDF. Finite size
estimates of the LDF,
\begin{equation}\label{eq:LDFfiniteDef}
I_\rho(m;L) \equiv -\frac{1}{L} \log \,\text{Prob}\,(n_{\max} = m L),
\end{equation}
were measured for several systems sizes at various system
parameters, see \fig \ref{fig:NumericalLDF}. A scaling collapse is
found, indicating that the condensate occupation indeed satisfies a
large deviation principle. Furthermore, there is a good agreement
between the MF prediction (\ref{eq:LDFfinal}) and this scaling
collapse. Since $\rho_\text{trans}(a)$ is defined by this LDF (it is
the density where the two local minima are equal), if the latter is
correct to high accuracy so must the former be. Note also that the
minimum of the $m>a$ branch is located close to $\rho-1$, in
agreement with the predicted edge of stability line $\rho = a+1$.
One caveat is that, as mentioned above, the background density
$\rho_\infty$ in the condensed phase is seen to deviate slightly
from the MF value 1. This deviation might indicate that the MF
prediction for $I_\rho(m)$ is not exact after all, since the $m>a$
minimum of this LDF must be located at $m=\rho-\rho_\infty$.

\subsection{Dynamics of the condensate}

\begin{figure*}
        \centering
        \includegraphics[width=0.4\textwidth]{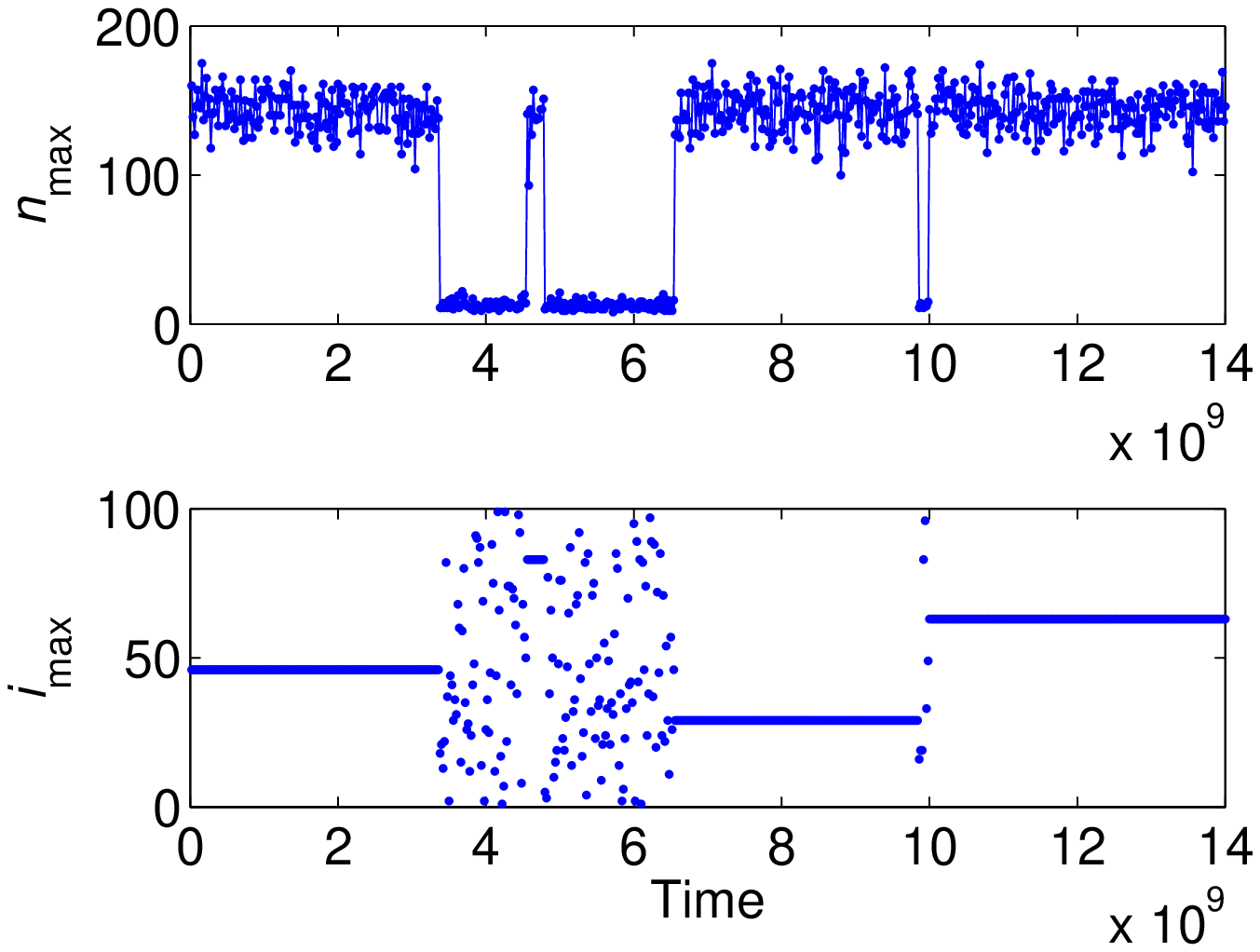}
        \includegraphics[width=0.4\textwidth]{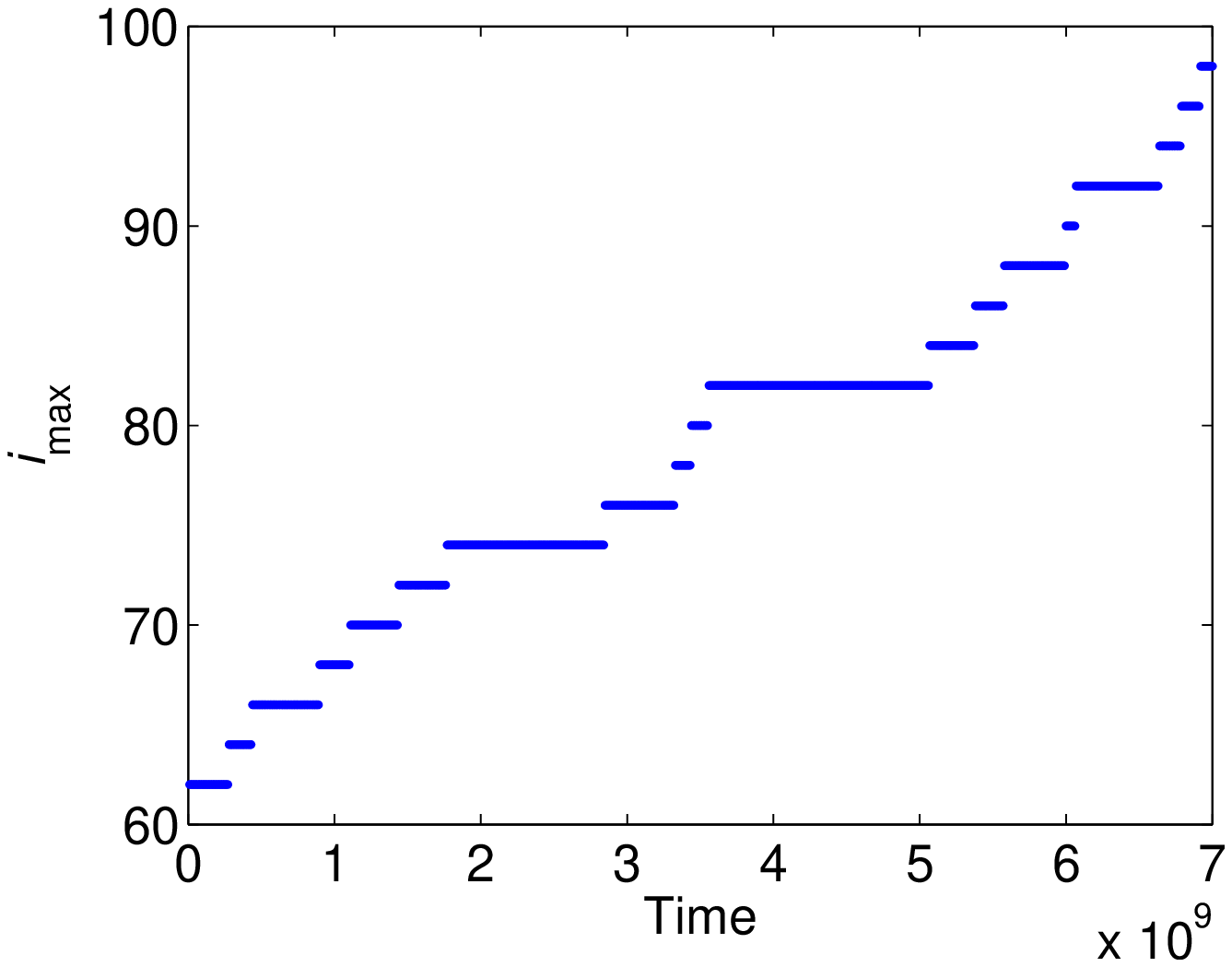}
        \caption{The dynamics of the condensate. Left panel: slightly above
        the transition density, the condensate typically relocates to
        a random site by
        switching between the condensed and disordered metastable
        states. Here we plot the occupation (top) and location (bottom)
        of the most occupied site as a function of time, in a system
        with $L=100$, $a = 0.5$, and $\rho = 2.44$. Note that this
        density is somewhat below $\rho_\text{trans}(a)$, but it is
        above its finite-size counterparts $\rho_\text{trans}(a;L)$,
        see Appendix \ref{sec:AppendixNumericalPhaseDiagram}. Right
        panel: at higher densities, the condensate typically
        relocates to its next nearest neighbor [the occupation $n_{\max}$,
        not plotted here, fluctuates around $(\rho-1)L$]. Here $L=160$, $a=0.15$,
        and $\rho = 4$. Note the scale of the time axes.
        \label{fig:NumericalCondDrift} }
\end{figure*}

Finally, we study the dynamics of the condensate when $\lmax = a L$.
This is done by recording how the location $i_{\max}$ and occupation
$n_{\max}$ of the most occupied site evolve with time. As discussed
above, the condensate dynamics is dictated by two competing
processes: switching to the disordered metastable state, and
condensate relocation to another site. At densities close to
$\rho_{\text{trans}}(a)$ switching is much faster and thus
dominates, while at higher densities switching becomes slower and
condensate relocation dominates. Measurements of the dynamics in the
two regimes are displayed in \fig \ref{fig:NumericalCondDrift}. When
the condensate relocates it is always seen to reappear in its
next-nearest neighbor site thus leading to a condensate drift, as
predicted by the MF theory.

\section{Summary and outlook}\label{sec:Conclusion}

In this paper we have presented a mean-field analysis of the AEP.
According to this analysis, a condensation transition takes place in
the model, but only when $\lmax$ increases at least logarithmically
with the system size. The critical density and current which are
found in this case are in agreement with the numerical results
presented here and in Ref.\ \cite{DongAEPprl2012}. Unlike naive MF
approximations, the MF scheme employed here (following Ref.\
\cite{CondDriftPRE2013}) takes into account some of the effects of
spatial correlations in the steady state. We thus find that in the
condensed phase, the fluid background is not homogeneous: the mean
occupation of a site and its fluctuations depend on the distance
from the condensate. As a result, there is, in the thermodynamic
limit, an overwhelming probability that when the condensate
relocates it does so to its next-nearest neighbor downstream.
Therefore, in the slow timescale of the condensate dynamics, it
drifts steadily along the lattice. Note that the drift velocity of
the condensate in the AEP decays at least algebraically with the
system size $L$, and in particular when $\lmax = a L$ it decays
exponentially with $L$. This is a much faster decay than the
$L^{-1}$ decay of the drift velocity in the model analyzed in Refs.\
\cite{HirschbergEtal2009,*HirschbergEtal2012Long,CondDriftPRE2013}.

In the analysis presented above we have made two assumptions that we
are not able to justify a-priori. The central one is the MF
approximation, which, as is usually the case, is uncontrolled. A
second assumption, made during the calculation of the condensate LDF
when $\lmax = a L$, is that the condensate evolution is adiabatic,
in the sense that the fluid background is always in its steady state
with the momentary condensate occupation (see \sect
\ref{sec:CondensateLDFcalculation}). To verify the validity of our
results, we have compared them with numerical simulations of the
model, and have found a good qualitative agreement between the
theoretical predictions and numerics. Furthermore, the numerical
results agree also quantitatively to a high degree with the
predicted MF critical density and current when $\lmax \ll L$, and
with the MF phase diagram when $\lmax = a L$. It is not yet known
whether these MF predictions are exact. Analyzing the model beyond
MF approximations is an interesting, and undoubtfully difficult,
open problem.

The current work joins Refs.\
\cite{HirschbergEtal2009,*HirschbergEtal2012Long,CondDriftPRE2013}
in demonstrating that a condensate drift is a rather generic
phenomenon in spatially-correlated nonequilibrium systems where a
condensate spontaneously breaks the translation symmetry. It would
be interesting to study from a more general perspective the
symmetry-breaking aspect of the drift mechanism. Recently, a similar
emergent motion was found in a system where a different symmetry is
spontaneously broken: a phase with a rotating magnetization was
established in a driven $XY$ model
\cite{MaesShlosman2011RotatingXYmodel}. It would be very interesting
to find other systems that exhibit a motion induced by the
combination of spontaneous symmetry breaking and a drive, and to
ascertain how general this phenomenon is.


\begin{acknowledgments}
We thank A.\ Bar, O.\ Cohen, T.\ Sadhu, and R.\ K.\ P.\ Zia for
useful discussions. The support of the Israel Science Foundation
(ISF) and of the Minerva Foundation with funding from the Federal
German Ministry for Education and Research is gratefully
acknowledged.
\end{acknowledgments}

\appendix

\section{Bulk properties in the condensed phase when $\lmax \ll L$}
\label{sec:AppendixFib}

In this appendix, we complete the details of the calculation of
$P_i(0)$, $\q_i$, and $\Q_i$ in the condensed phase. Assuming that
no site other than the condensate has more than $\lmax$ particles,
one obtains \eqn (\ref{eq:AEPp0Recursion}) from \eqns
(\ref{eq:AEPqDefinition}) and (\ref{eq:AEPp0}). As shown in \sect
\ref{sec:SecondOrderSupercritical}, the boundary condition for this
recursion relation is $P_2(0) = 1/2$ and $P_3(0)=1/3$.

Calculating the next few terms in the sequence yields
\begin{equation}
P_4(0) = \frac{2}{5}, \qquad P_5(0) = \frac{3}{8},
\quad \text{and} \quad P_6(0) = \frac{5}{13},
\end{equation}
leading us to guess $P_i(0) = F_{i-1}/F_{i+1}$, where $F_i$ is the
$i$'th Fibonacci number. We prove this guess by induction:
substituting our guess in the right hand side of \eqn
(\ref{eq:AEPp0Recursion}) yields
\begin{multline}
P_{i+1}(0) = \frac{1-[1-P_i(0)][1- P_{i-1}(0)]}{2- P_i(0)} = \\
\frac{1-\bigl[\frac{F_i}{F_{i+1}}\bigr]\,
\bigl[\frac{F_{i-1}}{F_{i}}\bigr]}{2-\frac{F_{i-1}}{F_{i+1}}}  =
\frac{\frac{F_{i+1}-F_{i-1}}{F_{i+1}}} {\frac{F_{i+1} +
(F_{i+1}-F_{i-1})}{F_{i+1}}} = \frac{F_i}{F_{i+2}},
\end{multline}
where we have repeatedly used $F_{i+1} - F_{i-1} = F_i$.

Finally, substituting in (\ref{eq:AEPqDefinition}) yields $q_i =
F_{i-2}/F_i$ and $Q_i = 1$.

\section{Fixed point analysis of \eqn (\ref{eq:AEPp0Recursion})}
\label{sec:AppendixFixedPoint}

In this Appendix, we analyze the fixed points of the recursion
relation for $P_i(0)$, \eqn (\ref{eq:AEPp0Recursion}). To simplify
notation, we denote in this Appendix
\begin{equation}
p_i \equiv P_i(0).
\end{equation}
First, we rewrite the second order recursion relation
(\ref{eq:AEPp0Recursion}) as a two-dimensional first order one
\begin{equation}\label{eq:FPmap}
(p_{i+1},p_i) = f(p_i,p_{i-1}), \; \text{with} \;
f(x,y) = \Bigl(y,\frac{x+y-xy}{2-y}\Bigr).
\end{equation}
The map (\ref{eq:FPmap}) has a one-parameter family of fixed points,
the line $x=y$. Therefore, any value of $p_i$ is a fixed point of
the map.

Denote by $d(x,y) \equiv (y-x)/\sqrt{2}$ the signed distance of the
point $(x,y)$ from the fixed point line. By signed distance we mean
that $|d(x,y)|$ is the Euclidian distance, and $d$ is positive if
the point is above the line ($y>x$) and negative if it is below it
($y<x$). A simple calculation shows that
\begin{equation}
d\bigl(f(x,y)\bigr) = -\frac{1-y}{2-y}\,\frac{y-x}{\sqrt{2}} = -\frac{1-y}{2-y}\, d(x,y).
\end{equation}
Since $0\leq (1-y)/(2-y)\leq 1/2$ for all physical values $0\leq y
\leq1$, the distance shrinks by more than a factor of $1/2$ with
each iteration of the map. Therefore, the map is an exponential
contraction, in the sense that any initial point $(x,y)$ with $0\leq
x,y \leq 1$ approaches the line $x=y$ exponentially rapidly. It is
also seen that consecutive iterations oscillate between the two
sides of the fixed point map, i.e., $P_i(0)$ decays to its limiting
value via oscillations.

\section{Equation for the LDF of the two most occupied sites}
\label{sec:Appendix2dLDF}

In this Appendix we discuss the equation for the LDF of the
occupation of the two most occupied sites. Denote
$P_{\text{max}}^{(2)}(mL, m_2 L) \equiv P(n_{\max} = mL,
n_{\max}^{(2)} = m_2 L)$, where $n_{\max}$ is the occupation of the
most occupied site and $n_{\max}^{(2)}$ is the second most occupied
site. The probability $P_{\text{max}}^{(2)}$ satisfies a master
equation similar to (\ref{eq:LDFmasterEq}). Substituting the LDF
ansatz
\begin{equation}
P_{\text{max}}^{(2)}(mL, m_2 L) \sim e^{-LI_\rho^{(2)}(m,m_2)},
\end{equation}
keeping the leading order in $L$ and examining the steady state
leads to
\begin{align}\label{eq:LDF2DwkbEq}
0 &=\bigl(1-e^{\partial_1 I_{\rho}^{(2)}(m,m_2)}\bigr)
\bigl[e^{-\partial_1 I_{\rho}^{(2)}(m,m_2)}-\jin(m,m_2)\bigr] + {} \nonumber \\
{} &\phantom{{}={}} \bigl(1-e^{\partial_2 I_{\rho}^{(2)}(m,m_2)}\bigr)
\bigl[e^{-\partial_2 I_{\rho}^{(2)}(m,m_2)}-\jin^{(2)}(m,m_2)\bigr],
\end{align}
where $\partial_1 = \partial/\partial m$, $\partial_2 =
\partial/\partial m_2$, $\jin$ is the current into the most occupied
site, and $\jin^{(2)}$ is the current into the second most occupied
site \footnote{The explicit forms of $\jin$ and $\jin^{(2)}$ will
not be important for us and so we do not write them here.
$\jin(m,m_2)$ is similar to (\ref{eq:LDFjin}) with $\rhobg = \rho -
m - m_2$; $\jin^{(2)}(m,m_2)$ also has a similar form, but with
$\q_\infty$ and $\Q_\infty$ replaced by $\q_3$ and $\Q_3$}. In \eqn
(\ref{eq:LDF2DwkbEq}) we have already used the fact that the current
out of both highly occupied sites condensate is 1. This equation
must be supplemented by boundary conditions, which are derived from
the boundary conditions of the master equation for $P_{\max}^{(2)}$.
In particular, when $m_2 = 0$, one finds that $\tilde{I}_\rho(m)
\equiv I_\rho^{(2)}(m,0)$ satisfies \eqn (\ref{eq:LDFwkbEq}) with
(\ref{eq:LDFjin}) and (\ref{eq:LDFrhobgApprox}).

A simple solution of the PDE (\ref{eq:LDF2DwkbEq}) may be found if
each of the two terms in the square brackets vanishes independently.
A necessary and sufficient condition for this to occur (found by
equating the mixed second derivatives of $I_\rho^{(2)}$) is that
\begin{equation}
\partial_2 \log \jin(m,m_2) = \partial_1 \log \jin^{(2)}(m,m_2).
\end{equation}
Unfortunately, a straightforward calculation shows that this is not
the case for the AEP.

\section{Numerical determination of the transition
density}\label{sec:AppendixNumericalPhaseDiagram}

\begin{figure}
        \centering
        \includegraphics[width=0.4\textwidth]{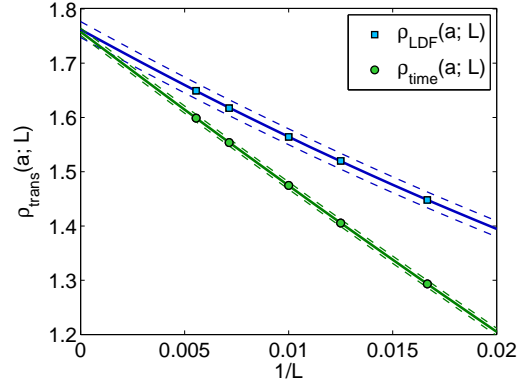}
        \caption{The transition density $\rho_\text{trans}(a)$ is estimated by
        extrapolating the finite-size estimators $\rho_\text{LDF}(a;L)$ and
        $\rho_\text{time}(a;L)$ to $L = \infty$. Markers are simulation
        measurements and lines are the best fits to a quadratic polynomial
        in $1/L$. Dashed lines mark the 95\% confidence intervals of the fits.
        The two different estimators agree to within measurement accuracy.
        \label{fig:NumericalRhoTransEstimation} }
\end{figure}

To numerically estimate the transition density
$\rho_\text{trans}(a)$ of \fig \ref{fig:PhaseDiagram} from
Monte-Carlo simulation data, we have used a finite-size scaling
analysis. In this Appendix we explain our calculational procedure.
The analysis is based on measuring finite-size estimators of the
critical density, and extrapolating these measurements to
$L=\infty$.

We study two natural finite-size estimators for
$\rho_\text{trans}(a)$.
\begin{enumerate}
  \item[(i)] We define $\rho_\text{LDF}(a;L)$ as the density at
      which the two local minima of the finite-size LDF
      $I_\rho(m;L)$ [\eqn (\ref{eq:LDFfiniteDef})] have the same
      value (see \fig \ref{fig:NumericalLDF}). Denote the
      location of the left minimum for a given values of $L$ and
      $a$ by $m_l^*$ and of the right by $m_r^*$. Then
      $\rho_\text{LDF}(a;L)$ is defined by
      \begin{equation}\label{eq:AppendixNumericalRhoLDFdef}
      I_{\rho_\text{LDF}(a;L)}(m_l^*;L) = I_{\rho_\text{LDF}(a;L)}(m_r^*;L).
      \end{equation}
      Note that for a given value of $L$ the density $\rho$ can
      only change by multiples of $1/L$, and thus there may not
      by a ``legitimate'' density at which this equality holds.
      In this case, we interpolate results to densities which
      are not multiples of $1/L$. Thus, $\rho_\text{LDF}(a;L)$
      need not be a multiple of $1/L$.
  \item[(ii)] As an alternative finite-size estimator, we define
      $\rho_\text{time}(a;L)$ as the density at which the system
      spends an equal fraction of the time in each of the two
      metastable states. To this end we measure the fraction of
      time $p_\text{dis}(\rho)$ that the system is in the
      disordered state, i.e., $p_\text{dis}(\rho) \equiv
      \text{Prob}(n_{\max} \leq \lmax)$. Then,
      $\rho_\text{time}(a;L)$ is defined by
      \begin{equation}
      p_\text{dis}\bigl(\rho_\text{time}(a;L)\bigr) = 1/2.
      \end{equation}
      Here too we interpolate $p_\text{dis}(\rho)$ to densities
      which are not multiples of $1/L$.
\end{enumerate}
Both of these estimators converge to $\rho_\text{trans}(a)$ when
${L\to\infty}$. At large system sizes, the second estimator can be
measured to a somewhat higher accuracy, because at the density
$\rho_\text{LDF}(a;L)$ the systems spends a small fraction of the
time in the disordered state, and thus more statistics (i.e., longer
simulation times) are needed in order to estimate $I_\rho(a;L)$ to a
high enough accuracy.

Next, we assume that both estimators can be expanded around
$L=\infty$ as $\rho_\text{trans}(a;L) = \rho_\text{trans}(a) + c_1/L
+ c_2/L^2 +\ldots$ for some coefficients $c_i$, where
$\rho_\text{trans}(a;L)$ stands for either of the two estimators. We
then fit the measured values to a quadratic function in $1/L$, and
extract the transition density, see \fig
\ref{fig:NumericalRhoTransEstimation}. We estimate the error as the
95\% confidence intervals of the fit; this in fact somewhat
underestimates the error, as it does not take into account the
statistical and systematic errors on the estimation of
$\rho_\text{trans}(a;L)$ (the latter are due to the interpolation
discussed above).

\bibliographystyle{apsrev4-1}
\bibliography{cond_drift_aep_new}

\end{document}